\documentclass[aps,prd,preprintnumbers,groupedaddress,nofootinbib,amssymb,notitlepage,eqsecnum]{revtex4-2}
\usepackage{here}
\usepackage[dvipdfmx]{graphicx}
\usepackage{amsmath,amsthm,amssymb}
\usepackage{bm}
\usepackage{color}

\usepackage{amsfonts}
\usepackage{dcolumn}
\usepackage{hyperref}
\allowdisplaybreaks[1]
\usepackage{stackengine}

\newcommand{\be}{\begin{equation}}  
\newcommand{\ee}{\end{equation}}
\newcommand{\ba}{\begin{eqnarray}}
\newcommand{\ea}{\end{eqnarray}}

\newcommand{\bem}{\begin{bmatrix}}
\newcommand{\eem}{\end{bmatrix}}
\newcommand{\Mpl}{M_{\rm Pl}}

\allowdisplaybreaks

\begin{document}

\preprint{WUCG-26-01}

\title{Realizing the phantom-divide crossing with vector and scalar fields}

\author{Shinji Tsujikawa\footnote{{\tt tsujikawa@waseda.jp}}}

\affiliation{Department of Physics, Waseda University, 
3-4-1 Okubo, Shinjuku, Tokyo 169-8555, Japan}

\begin{abstract}

In generalized Proca theories, characterized by a vector field 
with broken $U(1)$ gauge invariance, late-time cosmic acceleration 
can be realized with a dark energy equation of state in the regime
$w_{\rm DE} < -1$.
In such scenarios, however, a phantom-divide crossing, as recently 
suggested by DESI observations, is not  achieved without
encountering theoretical inconsistencies. 
We incorporate a canonical scalar field with a potential,
in addition to the vector field, and show that the phantom-divide
crossing from $w_{\rm DE} < -1$ to $w_{\rm DE} > -1$
can occur at low redshifts. 
We propose a minimal model that admits such a transition and identify
the region of parameter space in which all dynamical degrees of
freedom in the scalar, vector, and tensor sectors are free from ghost
and Laplacian instabilities. 
We further investigate the evolution of linear cosmological
perturbations by applying the quasi-static
approximation to modes well inside the Hubble radius. 
The dimensionless quantities $\mu$ and $\Sigma$, which characterize
the growth of matter perturbations and the bending of light rays,
respectively, depend on the sound speed $c_\psi$ of the longitudinal
scalar perturbation associated with the vector field. 
Since $c_\psi$ is influenced by the transverse vector mode, the model
exhibits sufficient flexibility to yield values of $\mu$ and
$\Sigma$ close to 1. 
Consequently, unlike theories such as scalar Galileons, 
the present model can be consistent with observations of redshift-space
distortions and integrated Sachs-Wolfe-galaxy cross-correlations.

\end{abstract}

\date{\today}

%\pacs{04.50.Kd, 95.36.+x, 98.80.-k}

\maketitle

%%%%%%%%%%%%%%%%%%%%%%%%%%%%%%%%%
\section{Introduction}
\label{introsec}
%%%%%%%%%%%%%%%%%%%%%%%%%%%%%%%%%

The landmark discovery of the late-time cosmic acceleration in 1998 
\cite{SupernovaSearchTeam:1998fmf,SupernovaCosmologyProject:1998vns} 
triggered a broad observational effort to unveil the nature of 
dark energy (DE) (see Refs.~\cite{Sahni:1999gb,Carroll:2000fy,Peebles:2002gy,Padmanabhan:2002ji,Copeland:2006wr,Silvestri:2009hh,Clifton:2011jh,Tsujikawa:2013fta,Joyce:2014kja,Koyama:2015vza,Bull:2015stt,Heisenberg:2018vsk,Kase:2018aps} for reviews). 
Since then, successive cosmological probes have yielded increasingly precise 
constraints on its equation of state, $w_{\rm DE}$. 
Early analyses based on Type~Ia supernova (SN~Ia) data~\cite{SupernovaSearchTeam:1998fmf,SupernovaCosmologyProject:1998vns} adopted the simplest assumption that DE originates from a cosmological constant 
with $w_{\rm DE} = -1$. Independent confirmation of the existence of DE came from measurements of the Cosmic Microwave Background (CMB) temperature anisotropies by WMAP~\cite{WMAP:2003elm}, thereby raising the intriguing possibility that the origin of DE may not be a pure cosmological constant.
The subsequent detection of baryon acoustic oscillations (BAOs) 
in galaxy surveys \cite{SDSS:2005xqv}
further enriched the observational landscape, opening new avenues to probe 
the properties of DE.

Recent BAO measurements from the Dark Energy Spectroscopic Instrument (DESI)~\cite{DESI:2024mwx,DESI:2024aqx,DESI:2025zgx}
have revealed a notable preference for dynamical DE scenarios in which
the equation of state $w_{\rm DE}$ evolves with redshift $z$, 
in contrast to the cosmological constant~$\Lambda$.
Using the Chevallier-Polarski-Linder 
(CPL) parametrization \cite{Chevallier:2000qy,Linder:2002et},
$w_{\rm DE} = w_0 + w_a z/(1+z)$,
with $w_0$ and $w_a$ treated as free constants,
a joint statistical analysis of the DESI DR2 measurements \cite{DESI:2025zgx}
and the Planck~2018 CMB observations indicates that models 
with $w_a \neq 0$
are favored over the $\Lambda$-Cold-Dark-Matter ($\Lambda$CDM) model 
at the $3.1\sigma$ level. This preference for dynamical DE persists 
when SN~Ia data are included, with a statistical significance of at
least $2.8\sigma$~\cite{DESI:2025fii}.
These results suggest that a time-varying DE component
may provide a better description of the current expansion history
than a cosmological constant.

Intriguingly, the DESI data further hint at a crossing of the phantom divide,
from $w_{\rm DE}<-1$ to $w_{\rm DE}>-1$, at low redshifts 
$z\lesssim1$. In conventional dynamical DE models, such as quintessence \cite{Fujii:1982ms,Ratra:1987rm,Wetterich:1987fm,Chiba:1997ej,Ferreira:1997au,Caldwell:1997ii,Copeland:1997et} and k-essence \cite{Armendariz-Picon:1999hyi,Chiba:1999ka,Armendariz-Picon:2000nqq}, 
the DE equation of state is generically restricted to 
$w_{\rm DE}>-1$ to avoid ghost and Laplacian instabilities (see 
Refs.~\cite{Shlivko:2025fgv,Akrami:2025zlb,Bayat:2025xfr,Cline:2025sbt,Gialamas:2025pwv,Alestas:2025syk,Shlivko:2025krk} for DESI constraints on quintessence models).
Although an equation of state with $w_{\rm DE} < -1$ can be realized by
introducing a phantom scalar field with a negative kinetic term
\cite{Caldwell:1999ew,Caldwell:2003vq,Singh:2003vx}, this construction suffers from severe 
theoretical pathologies associated with vacuum instabilities
\cite{Carroll:2003st,Cline:2003gs}. 
Even in the quintom scenario, in which a canonical quintessence field is
combined with a phantom scalar field to enable a phantom-divide
crossing~\cite{Feng:2004ad,Guo:2004fq}, the presence of a ghost 
renders the vacuum catastrophically unstable.

If derivative self-interactions of scalar or vector fields are introduced, 
it is well known that a DE equation of state with $w_{\rm DE}< -1$ 
can be achieved without invoking ghost instabilities.
Cubic scalar Galileon theories~\cite{Nicolis:2008in,Deffayet:2009wt} provide 
a representative example, in which a tracker solution evolves from 
a phantom-like regime ($w_{\rm DE}< -1$) toward a de 
Sitter attractor ($w_{\rm DE}= -1$)~\cite{DeFelice:2010pv,DeFelice:2010nf,Nesseris:2010pc}.
A similar evolution of $w_{\rm DE}$ can also arise in generalized Proca (GP) theories, 
which are second-order vector-tensor theories with broken $U(1)$ gauge invariance~\cite{Heisenberg:2014rta,Tasinato:2014eka,Allys:2015sht,BeltranJimenez:2016rff,Allys:2016jaq}.
In these theories, however, the DE equation of state must remain in the 
regime, $w_{\rm DE}< -1$, to avoid the appearance of ghost degrees of freedom~\cite{DeFelice:2016yws,DeFelice:2016uil,deFelice:2017paw}.
Indeed, a no-go theorem forbidding a crossing of $w_{\rm DE}= -1$ holds in both shift-symmetric scalar-tensor theories and GP theories with a luminal speed of gravitational waves~\cite{Tsujikawa:2025wca}, 
based on a unified effective field theory description of 
the two classes of theories \cite{Aoki:2021wew,Aoki:2024ktc,Aoki:2025bmj}.

In Ref.~\cite{Tsujikawa:2025wca}, the breaking of shift symmetry is shown to 
play a crucial role in realizing a phantom-divide crossing.
For instance, a nonminimal coupling of the form $F(\phi)R$, where $\phi$ 
denotes a scalar field and $R$ is the Ricci scalar, explicitly violates the 
invariance under constant shifts $\phi \to \phi+c$.
It is well known that $f(R)$ models of cosmic acceleration, which form 
a subclass of nonminimally coupled theories, can accommodate 
a phantom-divide crossing from $w_{\rm DE}< -1$ to $w_{\rm DE}> -1$~\cite{Hu:2007nk,Starobinsky:2007hu,Appleby:2007vb,Tsujikawa:2007xu,Amendola:2007nt,Tsujikawa:2008uc,Linder:2009jz,Motohashi:2010tb}
(see also~\cite{Boisseau:2000pr,Capozziello:2002rd,Carroll:2003wy,Perivolaropoulos:2005yv,Amendola:2006kh,Amendola:2006we}).
However, viable $f(R)$ models must be carefully constructed 
to sufficiently suppress fifth forces in local regions of 
the Universe \cite{Hu:2007nk,Starobinsky:2007hu,Appleby:2007vb,Tsujikawa:2007xu,Faulkner:2006ub,Capozziello:2007eu}.
In particular, consistency with solar-system tests of gravity and measurements of cosmic 
structure growth requires the DE equation of state to remain close to $w_{\rm DE}=-1$ 
in $f(R)$ gravity~\cite{Hu:2007nk,Brax:2008hh}.

Another issue with nonminimally coupled DE is that, even if the propagation
of fifth forces is suppressed by screening mechanisms such as the 
Vainshtein mechanism \cite{Vainshtein:1972sx}, 
the gravitational coupling in overdense regions generally acquires 
a cosmological time dependence through the DE field. 
In such theories, the effective gravitational coupling $G_{\rm eff}$ is
proportional to $G_{\rm N}/F(\phi)$, where $G_{\rm N}$ denotes the Newton
gravitational constant~\cite{Babichev:2011iz,Kimura:2011dc}. 
The temporal variation of $G_{\rm eff}$ is tightly constrained by lunar laser ranging experiments~\cite{Hofmann:2018myc}, which yield 
$\dot{G}_{\rm eff}/(H_0 G_{\rm eff})=(0.99\pm1.06)\times10^{-3}\,(0.7/h)$, 
where $h$ is related to the present Hubble parameter through
$H_0 = 100\,h\,{\rm km\,s^{-1}\,Mpc^{-1}}$. 
These bounds place strong restrictions on the time variation of the DE
scalar field at low redshifts, suggesting that deviations of the DE
equation of state from $w_{\rm DE} = -1$ are required to be small
\cite{Tsujikawa:2019pih}. As a consequence, nonminimally coupled 
DE models recently proposed to realize a phantom-divide crossing~\cite{Ye:2024ywg,Wolf:2024stt,Ye:2024zpk,Pan:2025psn,Wolf:2025jed,Wang:2025znm,Adam:2025kve,SanchezLopez:2025uzw} are generically expected to be observationally indistinguishable from a cosmological constant. Closely related scenarios include coupled DE and DM models, which can be
interpreted as Einstein-frame representations obtained via a conformal
transformation of nonminimally coupled theories.
By introducing an effective equation of state that combines both the DE
and interaction sectors, the authors of
Refs.~\cite{Chakraborty:2025syu,Khoury:2025txd,Guedezounme:2025wav}
argued that a phantom-divide crossing may occur.
Nevertheless, unless the equation of state of the pure DE component itself
crosses $w_{\rm DE} = -1$, it remains questionable whether such models can
be fully consistent with current observational data~\cite{Linder:2025zxb}.

A different mechanism for breaking the shift symmetry is to introduce a potential $V(\phi)$ for a canonical scalar field. 
In Ref.~\cite{Tsujikawa:2025wca}, such a potential is incorporated into the Lagrangian of the Galileon Ghost Condensate (GGC) model~\cite{Peirone:2019aua}, which belongs to a class of shift-symmetric 
Horndeski theories \cite{Horndeski:1974wa}. 
This framework allows for a phantom-divide crossing at low redshifts, from $w_{\rm DE}< -1$ to $w_{\rm DE}> -1$. Since nonminimal couplings are absent, issues related to the propagation of fifth forces or the time variation of the gravitational coupling in local regions of the Universe do not arise. The remaining challenge concerns the enhanced growth rate of matter density perturbations on scales relevant to large-scale structure. However, the large cosmic growth typically induced by Galileon interactions can be alleviated by the presence of the scalar potential \cite{Tsujikawa:2025wca}. A statistical analysis of a subclass of the GGC model with the potential $V(\phi)=V_0+m^2\phi^2/2$ indicates a strong preference over the cosmological constant, with a Bayes factor of 
$\ln B=6.5$~\cite{Wolf:2025acj}.

In this paper, we propose an alternative mechanism for realizing a
phantom-divide crossing within GP theories by introducing a canonical
scalar field endowed with a potential $V(\phi)$. The resulting theory
belongs to a subclass of scalar-vector-tensor theories, whose
cosmological dynamics have been extensively studied in
Refs.~\cite{Heisenberg:2018acv,Heisenberg:2018mxx,Kase:2018nwt,Heisenberg:2018wye}.
In this framework, the vector field with broken $U(1)$ gauge invariance
drives the DE equation of state into the phantom regime,
$w_{\rm DE}<-1$, without inducing ghost instabilities. The scalar
field, on the other hand, facilitates a subsequent transition to
$w_{\rm DE}>-1$ at low redshifts through its shift-symmetry-breaking
potential. By focusing on a subclass of scalar-vector-tensor theories 
without nonminimal couplings to either the Ricci scalar or the Einstein tensor, 
the model automatically satisfies observational constraints on the speed 
of gravitational waves, as well as local gravity tests.

Unlike the GGC model with a scalar potential discussed above---where 
$w_{\rm DE}$ becomes positive at high redshifts---the present scenario
predicts that $w_{\rm DE}$ asymptotically approaches $-1$ in the early Universe,
owing to the dominance of the scalar potential energy. The quantities $\mu$ and
$\Sigma$, which govern the growth of matter density perturbations and the
deflection of light rays \cite{Amendola:2007rr,Zhao:2010dz,Song:2010fg,Simpson:2012ra}, 
are controlled by the sound speed $c_{\psi}$ of the
longitudinal scalar mode arising from the vector field. In contrast to
scalar-tensor theories, $c_{\psi}$ is further affected by the transverse vector
mode, introducing an additional degree of freedom that allows the model to
remain compatible with observational constraints on the growth of large-scale
structure. We show that the phantom-divide crossing occurs within
regions of parameter space that are free from ghost and Laplacian instabilities.
Furthermore, we analyze the evolution of linear perturbations and explore key
observational signatures of the model, including redshift-space distortions and
integrated Sachs-Wolfe (ISW)-galaxy cross-correlations.

This paper is organized as follows.
In Sec.~\ref{backsec}, we present a concrete scalar-vector-tensor model that admits a phantom-divide crossing and obtain the background equations of motion in autonomous form.
In Sec.~\ref{linearsec}, we derive the second-order actions for tensor, vector, and scalar perturbations, and delineate the region of parameter space allowed by linear stability conditions.
In Sec.~\ref{dividesec}, we investigate how the phantom-divide crossing can occur at low redshifts due to the presence of the scalar potential.
In Sec.~\ref{growthsec}, we explore the observational signatures of the model associated with the evolution of linear perturbations, employing the quasi-static approximation for modes deep inside the Hubble radius.
Finally, Sec.~\ref{consec} is devoted to our conclusions. 
Throughout this paper, we adopt natural units: $c = 1$ and $\hbar = 1$.

%%%%%%%%%%%%%%%%%%%%%%%%%%%%%%%%%%%%%%%%%%%%%%
\section{Model and background equations of motion}
\label{backsec}
%%%%%%%%%%%%%%%%%%%%%%%%%%%%%%%%%%%%%%%%%%%%%%

We consider a vector field $A_\mu$ with the gauge-field 
strength $F = -F_{\mu\nu} F^{\mu\nu}/4$, 
where $F_{\mu\nu} = \partial_\mu A_\nu - \partial_\nu A_\mu$, 
together with a canonical scalar field $\phi$ with a potential $V(\phi)$. 
In addition to the term $F$, we introduce the Lagrangians 
$G_2(X)$ and $G_3(X) \nabla_{\mu}A^{\mu}$ in the vector-field sector, 
where $G_2$ and $G_3$ are functions of $X=-A_{\mu}A^{\mu}/2$, 
and $\nabla_{\mu}$ is a covariant derivative operator. 
The $U(1)$ gauge symmetry is explicitly broken 
by the $X$-dependent couplings in $G_2(X)$ and $G_3(X)$.
The action for this class of theories is given by
\be
{\cal S}=\int {\rm d}^4 x \sqrt{-g} \left[ \frac{\Mpl^2}{2}R+F
+G_2(X)+G_3(X) \nabla_{\mu}A^{\mu} 
-\frac{1}{2} g^{\mu \nu} \nabla_{\mu} \phi \nabla_{\nu} \phi 
-V(\phi) \right]+{\cal S}_{M}\,,
\label{action}
\ee
where $g$ is the determinant of the metric tensor $g_{\mu\nu}$,
$M_{\rm Pl}$ is the reduced Planck mass, $R$ is the Ricci scalar,
and $\mathcal{S}_M$ denotes the matter action. 
The first four terms inside the square brackets in Eq.~(\ref{action})
belong to a subclass of GP theories
\cite{Heisenberg:2014rta,Tasinato:2014eka,Allys:2015sht,BeltranJimenez:2016rff,Allys:2016jaq} that lead to second-order equations of motion.
In this case, the realization of the phantom-divide crossing
is prohibited without introducing theoretical 
pathologies \cite{Aoki:2024ktc,Tsujikawa:2025wca}.
As we will show in Sec.~\ref{dividesec}, this no-go theorem can be circumvented
by introducing a scalar-field potential $V(\phi)$, which explicitly
breaks the shift symmetry under the transformation $\phi \to \phi + c$. 
The key difference from the model proposed in Ref.~\cite{Tsujikawa:2025wca}
is that, in our case, the vector field plays a crucial role in realizing
a DE equation of state $w_{\rm DE} < -1$ prior to the phantom-divide crossing.

We study the background cosmological dynamics on the spatially flat
Friedmann-Lema\^{\i}tre-Robertson-Walker (FLRW) spacetime described by the
line element
\be
{\rm d}s^2=-{\rm d}t^2+a^2(t) \delta_{ij} 
{\rm d}x^i {\rm d}x^j\,,
\label{backmet}
\ee
where $a(t)$ is the scale factor and $t$ denotes the cosmic time.
On this background, the vector and scalar fields take the forms
\be
A^{\mu}=\left[ \chi(t),0,0,0 \right]\,,\qquad 
\phi=\phi(t)\,,
\ee
where $\chi$ and $\phi$ are the functions of $t$.
In the matter sector, we consider nonrelativistic matter with
energy density $\rho_m$ and vanishing pressure, together 
with radiation with energy density $\rho_r$ and pressure $P_r = \rho_r/3$.
These components satisfy the continuity equations
$\dot{\rho}_m + 3 H \rho_m = 0$
and
$\dot{\rho}_r + 4 H \rho_r = 0$,
respectively, where a dot denotes differentiation
with respect to cosmic time $t$. 
Varying the action~(\ref{action}) with respect to $g_{\mu\nu}$,
we obtain the gravitational equations of motion:
\ba
& &
3 M_{\rm pl}^2 H^2 = \rho_{\rm DE} + \rho_m + \rho_r, \label{back1}\\
& &
M_{\rm pl}^2 \left( 3H^2 + 2\dot{H} \right)
= - P_{\rm DE} - \frac{1}{3}\rho_r\,,\label{back2}
\ea
where $H \equiv \dot{a}/a$ is the Hubble expansion rate.
The DE density and pressure are given, respectively, by
\ba
\rho_{\rm DE} &=& -G_2+\frac{1}{2}\dot{\phi}^2+V(\phi)\,,\\
P_{\rm DE} &=& G_2-G_{3,X} \dot{\chi} \chi^2
+\frac{1}{2}\dot{\phi}^2-V(\phi)\,,
\ea
where we use the notation 
$G_{i,X} \equiv {\rm d} G_{i}/{\rm d}X$, with $i=2,3$. 
Defining the DE equation of state as
$w_{\rm DE} \equiv P_{\rm DE}/\rho_{\rm DE}$, we obtain
\be
w_{\rm DE}=-1+\frac{G_{3,X} \dot{\chi}\chi^2-\dot{\phi}^2}
{G_2-\dot{\phi}^2/2-V}\,.
\label{wDE0}
\ee
This result shows that the time evolution of $\chi$ and $\phi$
leads to deviations of $w_{\rm DE}$ from $-1$.
The vector and scalar fields satisfy the following equations
of motion:
\ba
& &
\chi \left( G_{2,X}+3H \chi\,G_{3,X} \right)=0\,,
\label{back3}\\
& & 
\ddot{\phi}+3H \dot{\phi}+V_{,\phi}=0\,,
\label{back4}
\ea
where $V_{,\phi} \equiv {\rm d}V/{\rm d}\phi$.
Equation~(\ref{back3}) admits two branches:
(i) $\chi = 0$ and
(ii) $G_{2,X} + 3 H \chi G_{3,X} = 0$.
In case (i), the model~(\ref{action}) reduces to a quintessence
scenario driven by the potential $V(\phi)$, and hence the
phantom-divide crossing cannot be realized.
In case (ii), the temporal component of the vector field $\chi$
is generically nonvanishing.
Since the couplings $G_2$ and $G_3$ are functions of
$X = \chi^2/2$ in this branch, $\chi$ can be expressed
as a function of $H$.
In the following, we focus on case (ii), which permits 
the phantom-divide crossing.

For concreteness, we consider the coupling functions given by 
\be
G_{2}(X) = b_{2} X^{p_{2}}\,,\qquad
G_{3}(X) = b_{3} X^{p_{3}}\,,
\label{G23}
\ee
where $b_2, b_3, p_2, p_3$ are constants. 
From Eq.~(\ref{back3}), the field $\chi$ is related to $H$ as
\be
\chi^p H =-\frac{2^{p_{3}-p_{2}} b_{2}p_2}
{3 b_{3}p_{3}}\,,
\label{chipre}
\ee
where 
\be
p \equiv 1-2p_2+2p_3\,.
\ee
As long as $b_3 p_3 \neq 0$, a solution of the form
$\chi^{p} H = {\rm constant}$ exists.
Moreover, the product $\chi^{p} H$ is nonvanishing 
provided that $b_2 p_2 \neq 0$.
Therefore, both couplings $G_2(X)$ and $G_3(X)$ are required for
the existence of a solution of the type~(\ref{chipre}).
Without loss of generality, we focus on the branch with
$\chi > 0$. Provided that 
\be
p>0\,,
\label{pcon}
\ee
the field $\chi$ grows as $\chi \propto H^{-1/p}$ as $H$ decreases.
In this case, the energy density associated with the vector field, 
$\rho_{\chi}=-G_2=-b_2X^{p_2}=-b_2 (\chi^2/2)^{p_2}$,  
contributes to the DE density at late times. 
Since we are interested in the case $\rho_{\chi}>0$, 
i.e., $b_2<0$, we parameterize
\be
b_2 \equiv -m^2 \Mpl^{2(1-p_2)}\,,
\ee
where $m$ is a positive constant with dimensions of mass. 
With this choice, the energy density of the vector field becomes
$\rho_{\chi}=m^2 \Mpl^2 [\chi^2/(2\Mpl^2)]^{p_2}$.
We also introduce the following 
dimensionless quantities:
\be
\nu \equiv u^p \frac{H}{m}\,,\qquad 
u \equiv \frac{\chi}{\Mpl}\,.
\ee
Since we consider the solution with $\chi > 0$ in an expanding Universe,
it follows that $u > 0$ and $\nu > 0$. 
{}From Eq.~(\ref{chipre}), $\nu$ is constant in time.
The density parameter associated with the field $\chi$ is given by 
\be
\Omega_{\chi} \equiv \frac{\rho_{\chi}}{3\Mpl^2 H^2}
=\frac{2^{-p_2}m^2 u^{2p_2}}{3H^2}\,.
\label{Omechi}
\ee
In the scalar-field sector, we also introduce 
the dimensionless quantities:
\be
x  \equiv \frac{\dot{\phi}}{\sqrt{6}\Mpl H}\,,\qquad 
y \equiv \frac{\sqrt{V}}{\sqrt{3}\Mpl H}\,,
\label{xy}
\ee
together with the associated density parameter
$\Omega_{\phi}=x^2+y^2$.
Then, from the Hamiltonian constraint~(\ref{back1}), 
we obtain
\be
\Omega_m=1-\Omega_{\chi}-x^2-y^2-\Omega_{r}\,,
\label{Omem}
\ee
where $\Omega_m \equiv \rho_m/(3\Mpl^2 H^2)$ and 
$\Omega_r \equiv \rho_r/(3\Mpl^2 H^2)$. 
By using the relation~(\ref{chipre}), the term $G_{3,X}\dot{\chi}\chi^2$
appearing in $P_{\rm DE}$ can be rewritten as 
$G_{3,X} \dot{\chi}\chi^2=-2\Mpl^2 s \dot{H} \Omega_{\chi}$, where 
\be
s \equiv \frac{p_2}{p}=\frac{p_2}{1-2p_2+2p_3}\,.
\ee
Combining Eqs.~(\ref{back1}) and (\ref{back2}), we obtain
\be
h \equiv \frac{\dot{H}}{H^2}=-\frac{3+3x^2-3y^2-3\Omega_{\chi}+\Omega_r}
{2(1+s \Omega_{\chi})} \,, 
\label{h}
\ee
where Eq.~(\ref{Omem}) has been used to eliminate $\Omega_m$.
Using the background Eqs.~(\ref{back4}) and~(\ref{chipre}),
the dimensionless variables $\Omega_\chi$, $x$, $y$, and $\Omega_r$
satisfy the following autonomous system: 
\ba
\Omega_{\chi}' &=& -2(s+1) h \Omega_{\chi}\,,\label{auto1}\\
x' &=& -3x+\frac{\sqrt{6}}{2} \lambda y^2-hx\,,\label{auto2}\\
y' &=& -\frac{\sqrt{6}}{2} \lambda xy -hy\,,\label{auto3}\\
\Omega_r' &=& -2(2+h) \Omega_r\,,\label{auto4}
\ea
where a prime denotes differentiation with respect to
$N \equiv \ln a$, and 
\be
\lambda \equiv -\frac{\Mpl V_{,\phi}}{V}\,.
\ee
In the following, we focus on the case in which $\lambda$
is a constant, corresponding to an exponential potential, 
\be
V(\phi)=V_0 e^{-\lambda \phi/\Mpl}\,,
\ee
where $V_0$ is a constant. 
At the background level, the model is characterized by two parameters,
$s$ and $\lambda$. The $\Lambda$CDM model corresponds to the limit
$s \to 0$ and $\lambda \to 0$, with $x \to 0$.
In this limit, integrating Eqs.~(\ref{auto1}) and~(\ref{auto3}) yields
$\Omega_\chi \propto H^{-2}$ and $y \propto H^{-1}$, respectively,
so that $u = {\rm constant}$ and $V = {\rm constant}$ from
Eqs.~(\ref{Omechi}) and~(\ref{xy}).
For fixed values of the model parameters $s$ and $\lambda$,
Eqs.~(\ref{auto1})-(\ref{auto4}) can be integrated to determine
$\Omega_\chi$, $x$, $y$, and $\Omega_r$ once the initial conditions 
are specified. 
The density parameter of nonrelativistic matter, 
$\Omega_m$, is then determined from Eq.~(\ref{Omem}). 
The density parameter of DE is given by 
\be
\Omega_{\rm DE}=\Omega_{\chi}
+x^2+y^2\,, 
\ee
while the DE equation of state~(\ref{wDE0}) can be expressed as
\be
w_{\rm DE}=-1+\frac{2(hs \Omega_{\chi}
+3x^2)}{3(\Omega_{\chi}+x^2+y^2)}\,.
\label{wde}
\ee
We also define the effective equation of state 
$w_{\rm eff}$ as 
\be
w_{\rm eff}=-1-\frac{2}{3}h\,.
\ee
The condition for late-time cosmic acceleration 
is $w_{\rm eff} < -1/3$.
Since stability against linear perturbations restricts
the allowed parameter space of the model, we discuss
this issue in Sec.~\ref{linearsec} before proceeding
to a detailed analysis of the background dynamics
in Sec.~\ref{dividesec}. 

%%%%%%%%%%%%%%%%%%%%%%%%%%%%%%%%%%%%%%
\section{Linear perturbations and stability conditions}
\label{linearsec}
%%%%%%%%%%%%%%%%%%%%%%%%%%%%%%%%%%%%%%

To derive the conditions for linear stability, we consider perturbations
around the spatially flat FLRW background~(\ref{backmet}).
The perturbed line element, including tensor, vector, and scalar modes,
is given by 
\be
{\rm d}s^2=-\left( 1+2\alpha \right){\rm d}t^2
+2\left( \partial_i B+V_i \right) {\rm d}t\,{\rm d}x^i
+a^2(t) \left( \delta_{ij}+h_{ij} \right){\rm d}x^i {\rm d}x^j\,,
\label{permet}
\ee
where $\alpha$ and $B$ are scalar perturbations, with the notation
$\partial_i B \equiv \partial B / \partial x^i$,
$V_i$ denotes the vector perturbation, and $h_{ij}$ represents the
tensor perturbation, with the spatial coordinates $(x^1,x^2,x^3)=(x,y,z)$.
The perturbed quantities depend on both $t$ and $x^i$. 
In the scalar and vector sectors, we adopt the flat gauge, in which
the scalar perturbations $\zeta$ and $E$, as well as the vector perturbation
$F_i$ in the three-dimensional spatial line element, vanish. 
The vector perturbation $V_i$ satisfies the transverse condition
$\partial^i V_i = 0$, whereas the tensor perturbation $h_{ij}$ obeys
the transverse and traceless conditions
$\partial^j h_{ij} = 0$ and ${h_i}^i = 0$. 

The scalar field $\phi$ is decomposed into background and 
perturbed parts as
\be
\phi=\tilde{\phi}(t)+\delta \phi (t,x^i)\,,
\ee
where, in what follows, we omit the tilde on background quantities.
The vector field $A^\mu$ consists of a temporal component $A^0$ and
spatial components $A^i$, where the latter contain only perturbed
contributions. We decompose $A^0$ and $A_i$ as
\be
A^0=\chi(t)+\delta A (t,x^i)\,,\qquad 
A_i=\partial_{i} \psi(t,x^i)+Z_i(t,x^i)\,,
\ee
where $\delta A$ and $\psi$ are scalar perturbations, while
$Z_i$ is the vector perturbation satisfying the transverse
condition $\partial^i Z_i = 0$. 

To describe a perfect fluid in the matter sector,
we consider the Schutz-Sorkin 
action \cite{Schutz:1977df,Brown:1992kc,DeFelice:2009bx}
\be
{\cal S}_M=-\int {\rm d}^4 x 
\left[ \sqrt{-g}\,\rho_{M}(n)+J^{\mu} \left( \partial_{\mu} \ell
+{\cal A}_1 \partial_{\mu} {\cal B}_1
+{\cal A}_2 \partial_{\mu} {\cal B}_2 \right) \right]\,,
\label{SM}
\ee
where the energy density $\rho_M$ is a function of the fluid
number density $n$. 
The scalar quantity $\ell$ is a Lagrange multiplier, 
whereas the current $J^\mu$ is related to the fluid 
number density $n$ via $n = \sqrt{J^\mu J_\mu / g}$.
The temporal and spatial components of $J^\mu$ can be decomposed as
\be
J^0={\cal N}_0+\delta J\,,\qquad
J^k=\frac{1}{a^2(t)} \delta^{ki} \left( 
\partial_i \delta j+W_i \right)\,,
\ee
where $\mathcal{N}_0 = n_0 a^3$ is the conserved background fluid number,
with $n_0$ denoting the background value of $n$, and
$\delta J$, $\delta j$, and $W_i$ represent perturbations. 
Since $W_i$ satisfies the transverse condition
$\partial^i W_i = 0$, we can choose its components as
$W_i = [W_1(t,z), W_2(t,z), 0]$.
We adopt the same configuration for $V_i$ and $Z_i$, 
namely $V_i = [V_1(t,z), V_2(t,z), 0]$ and
$Z_i = [Z_1(t,z), Z_2(t,z), 0]$.
The Lagrange multiplier fields ${\cal A}_1$, ${\cal A}_2$, ${\cal B}_1$,
and ${\cal B}_2$ describe the vector modes in the matter sector.
Without loss of generality, we choose these fields 
in the forms \cite{DeFelice:2009bx,DeFelice:2016yws,DeFelice:2016uil}
\be
{\cal A}_1=\delta {\cal A}_1 (t,z)\,,\qquad 
{\cal A}_2=\delta {\cal A}_2 (t,z)\,,\qquad 
{\cal B}_1=x+\delta {\cal B}_1(t,z)\,,\qquad 
{\cal B}_2=y+\delta {\cal B}_2(t,z)\,,
\ee
where the perturbations $\delta {\cal A}_1$, 
$\delta {\cal A}_2$, $\delta {\cal B}_1$, and  
$\delta {\cal B}_1$ depend on $t$ and $z$.

The four-velocity $u_\mu$, normalized as 
$u_\mu u^\mu = -1$, is related to $J_\mu$ via
$u_\mu = J_\mu / (n \sqrt{-g})$.
Varying the action~(\ref{SM}) with respect 
to $J^\mu$, we obtain
\be
u_{\mu}=\frac{1}{\rho_{M,n}} \left( \partial_{\mu} \ell
+{\cal A}_1 \partial_{\mu} {\cal B}_1
+{\cal A}_2 \partial_{\mu} {\cal B}_2 \right)\,,
\label{umu}
\ee
where $\rho_{M,n} \equiv {\rm d}\rho_M/{\rm d} n$.
We write the Lagrange multiplier field $\ell$ in the form
\be
\ell=-\int^t \rho_{M,n} (\tilde{t})\,{\rm d} \tilde{t} 
-\rho_{M,n} v\,,
\label{ell}
\ee
where $v$ corresponds to the velocity potential. 
Indeed, after substituting Eq.~(\ref{ell}) into Eq.~(\ref{umu}),
the spatial components of $u_\mu$ can be expressed as
\be
u_i=-\partial_i v+v_i\,.
\ee
The vector components $v_i$, which satisfy the transverse
condition $\partial^i v_i = 0$, are related to $\delta {\cal A}_i$ as
\be
\delta {\cal A}_i=\rho_{M,n}v_i\,.
\ee
Since the time derivative of the background matter density $\rho_M$ is
given by $\dot{\rho}_M = \rho_{M,n} \dot{n}_0$, the conservation equation
of the fluid number density, $\dot{n}_0 + 3 H n_0 = 0$, translates to
$\dot{\rho}_M + 3 H n_0 \rho_{M,n} = 0$.
From the continuity equation $\dot{\rho}_M + 3 H (\rho_M + P_M) = 0$, 
where $P_M$ is the matter pressure,  
we then obtain the relation $\rho_{M,n} = (\rho_M + P_M)/n_0$, 
and hence $\delta \mathcal{A}_i = [(\rho_M + P_M)/n_0] v_i$.

After expanding the total action~(\ref{action}) up to quadratic order, 
including the Schutz-Sorkin action~\eqref{SM}, the resulting 
second-order action can be decomposed into contributions from 
the tensor, vector, and scalar sectors. 
In the following, we study the propagation of each sector in turn.

\subsection{Tensor sector}

For the tensor perturbation $h_{ij}$, we choose the configuration
$h_{11} = h_1(t,z)$, $h_{22} = -h_1(t,z)$, and $h_{12} = h_{21} = h_2(t,z)$,
which satisfies the transverse and traceless conditions. 
The second-order action of the tensor modes arising from the matter sector
is given by $({\cal S}_M^{(2)})_t=-\int {\rm d}t\,{\rm d}^3 x 
\sum_{i=1}^2 (1/2)a^3 P_M h_i^2$. 
Computing the other quadratic-order contributions in ${\cal S}$ and 
using the background Eq.~(\ref{back2}), the terms proportional to 
$h_i^2$ identically vanish. 
The resulting second-order action takes the form
\be
{\cal S}_t^{(2)}=\int {\rm d}t\,{\rm d}^3 x \sum_{i=1}^{2} 
\frac{a^3}{4} q_t \left[ \dot{h}_i^2-\frac{c_t^2}{a^2} 
(\partial h_i)^2 \right]\,,
\ee
where 
\be
q_t=\Mpl^2\,,\qquad c_t^2=1\,.
\ee
This shows that there are two propagating degrees of freedom in the tensor 
sector. Since $q_t > 0$, the theory is free of ghosts, and the tensor modes 
propagate at the speed of light. 
Therefore, the model is consistent with the gravitational-wave constraints
inferred from the GW170817 event \cite{LIGOScientific:2017zic} and 
its electromagnetic counterpart \cite{Goldstein:2017mmi}.

\subsection{Vector sector}

Expanding the matter action ${\cal S}_M$ up to quadratic order in vector 
perturbations, we obtain the second-order action $({\cal S}_M^{(2)})_v$ 
containing the four fields $W_i$, $\delta {\cal A}_i$, $\delta {\cal B}_i$, 
and $V_i$ \cite{DeFelice:2016yws,DeFelice:2016uil,Kase:2018nwt}. 
Varying $({\cal S}_M^{(2)})_v$ with respect to $W_i$, $\delta {\cal A}_i$, 
and $\delta {\cal B}_i$, we find that 
$W_i={\cal N}_0 (v_i-V_i)$, $v_i=V_i-a^2 \dot{\delta {\cal B}}_i$, 
and $\delta {\cal A}_i=[(\rho_M+P_M)/n_0]v_i=C_i$, where $C_i$ 
are time-independent constants. 
Integrating out the fields $W_i$ and $\delta{\cal A}_i$ from 
the action $({\cal S}_M^{(2)})_v$ then yields
\be
({\cal S}_M^{(2)})_v=\int {\rm d}t\,{\rm d}^3 x \sum_{i=1}^2 
\frac{a}{2} \left[ (\rho_M+P_M)v_i^2-\rho_M V_i^2  \right]\,.
\ee
Computing the remaining quadratic-order contributions to ${\cal S}$ 
and using the background equations of motion, we obtain the 
second-order action for vector perturbations in the form
\be
{\cal S}_v^{(2)}=\int {\rm d}t\,{\rm d}^3 x \sum_{i=1}^{2} 
\left[  \frac{a}{2} \dot{Z}_i^2-\frac{1}{2a} (\partial Z_i)^2 
-\frac{a}{2} \alpha_v Z_i^2+\frac{\Mpl^2}{4a} 
(\partial V_i)^2+\frac{a}{2} (\rho_M+P_M)v_i^2 \right]\,,
\label{Sv}
\ee
where 
\be
\alpha_v \equiv G_{2,X}+\left( \dot{\chi}+3H \chi 
\right)G_{3,X}\,.
\ee
Varying the action (\ref{Sv}) with respect to $V_i$, it follows that 
\be
\Mpl^2 \partial^2 V_i=2a^2 n_0 C_i\,,
\ee
where we used the property $[(\rho_M+P_M)/n_0]v_i=C_i$. 
In the small-scale limit, the perturbation $V_i$ is suppressed, 
such that the last two terms in Eq.~(\ref{Sv}) can be neglected 
relative to the others. 
Therefore, the second-order action~(\ref{Sv}) approximately reduces to
\be
{\cal S}_v^{(2)} \simeq \int {\rm d}t\,{\rm d}^3 x \sum_{i=1}^{2} 
\frac{a}{2} q_v \left[  \dot{Z}_i^2-\frac{c_v^2}{a^2} (\partial Z_i)^2 
-\frac{\alpha_v}{q_v} Z_i^2 \right]\,,
\label{Sv2}
\ee
where 
\be
q_v=1\,,\qquad c_v^2=1\,.
\ee
There are two transverse vector modes propagating 
at the speed of light, with the ghost-free conditions satisfied. 
Therefore, the linear stability of both vector and tensor perturbations 
does not impose any constraints on the model parameters.

\subsection{Scalar sector}

In the scalar sector, the matter action ${\cal S}_M$ contains three 
scalar perturbations, $\delta J$, $\delta j$, and $v$. 
Instead of $\delta J$, we define the matter density perturbation as
\be
\delta \rho_M=\frac{\rho_{M,n}}{a^3} \delta J
=\frac{\rho_M+P_M}{n_0 a^3} \delta J\,.
\ee
Then, at linear order, the perturbation in the fluid number density 
reduces to $\delta n = \delta \rho_M / \rho_{M,n}$.
Expanding ${\cal S}_M$ up to quadratic order in scalar perturbations, 
we obtain the second-order action $({\cal S}_M^{(2)})_s$ 
containing $\delta \rho_M$, $\delta j$, $v$, $\alpha$, $B$, 
and their derivatives \cite{DeFelice:2016yws,Kase:2018nwt,Kase:2018aps}. 
Varying this action with respect to $\delta j$ yields the relation 
$\partial \delta j=-a^3 n_0 (\partial v+\partial B)$.
After eliminating $\delta j$ from $({\cal S}_M^{(2)})_s$, we obtain 
the reduced matter action
\be
({\cal S}_M^{(2)})_s
=\int {\rm d}t\,{\rm d}^3x\,a^3 \left\{ \left( \dot{v}-3H c_M^2v-\alpha 
\right) \delta \rho_M -\frac{c_M^2 (\delta \rho_M)^2}
{2n_0 \rho_{M,n}}-\frac{n_0 \rho_{M,n}}{2a^2} 
\left[ (\partial v)^2+2\partial v\partial B \right]
+\frac{\rho_M}{2} \alpha^2-\frac{\rho_M}{2a^2} 
(\partial B)^2 \right\}\,,
\label{SMS2}
\ee
where $c_M^2$ denotes the matter sound speed squared, defined by
\be
c_M^2=\frac{P_{M,n}}{\rho_{M,n}}
=\frac{n_0 \rho_{M,nn}}{\rho_{M,n}}\,.
\ee
Taking into account the remaining quadratic-order contributions to ${\cal S}$ 
and using the background equations of motion, the second-order action for 
scalar perturbations can be written as
\be
{\cal S}_s^{(2)}=\int {\rm d}t\,{\rm d}^3 x\,{\cal L}_s\,,
\label{Ss}
\ee
where 
\ba
\hspace{-0.7cm}
{\cal L}_s &=& a^3 \biggl\{
\left(w_1\alpha+w_2\frac{\delta A}{\chi}\right)\frac{\partial^2 B}{a^2}
-w_3\frac{(\partial \alpha)^2}{a^2}-w_3\frac{(\partial \delta A)^2}
{4a^2 \chi^2}+w_4 \alpha^2+w_5\frac{\delta A^2}{\chi^2}
+\left[ w_3\dot{\psi}-(w_2+\chi w_6)\psi \right] 
\frac{\partial^2 \delta A}{2a^2 \chi^2}
\nonumber\\
\hspace{-0.7cm}
&&~~~\,+\left( w_3\frac{\partial^2\delta A}{a^2 \chi}
-w_8\frac{\delta A}{\chi}
+w_3\frac{\partial^2\dot{\psi}}{a^2 \chi}
-w_6\frac{\partial^2\psi}{a^2}\right)\alpha 
-w_3\frac{(\partial\dot{\psi})^2}{4a^2 \chi^2}+w_7\frac{(\partial\psi)^2}{2a^2}
\nonumber\\
\hspace{-0.7cm}
&&~~~\,
+\frac{1}{2} \dot{\delta \phi}^2-\frac{(\partial \delta \phi)^2}{2a^2}
-\frac{V_{,\phi \phi}}{2} \delta \phi^2 
-\left( \dot{\phi} \dot{\delta \phi}+V_{,\phi} \delta \phi \right)\alpha 
+\frac{\dot{\phi}}{a^2} \delta \phi\,\partial^2 B \nonumber\\
\hspace{-0.7cm}
&&~~~\,+\left( \rho_M+P_M \right)v\frac{\partial^2 B}
{a^2}-v\dot{\delta \rho}_M-3H (1+c_M^2) v\delta \rho_M 
-\frac{\rho_M+P_M}{2} \frac{(\partial v)^2}{a^2}
-\frac{c_M^2}{2 (\rho_M+P_M)} (\delta \rho_M)^2 
-\alpha \delta \rho_M \biggr\},
\label{Ls}
\ea
with
\ba
& &
w_1=-\chi^3 G_{3,X}-2\Mpl^2 H\,,\qquad 
w_2=-\chi^3 G_{3,X}\,,\qquad
w_3=-2\chi^2 q_v\,, \nonumber \\
& &
w_4=\frac{\chi^3}{2} \left[ \chi G_{2,XX}+3H 
(\chi^2 G_{3,X} -G_{3,X}) \right]+\frac{\dot{\phi}^2}{2} 
-3\Mpl^2 H^2\,,\qquad 
w_5=\frac{\chi^3}{2} \left[ \chi G_{2,XX}
+3H ( \chi^2G_{3,XX} +G_{3,X}) 
 \right]\,, \nonumber \\
 & &
 w_6=-\chi^2 G_{3,X}\,, \qquad
 w_7=-\dot{\chi}G_{3,X}\,,\qquad
 w_8=-\chi^4 \left( G_{2,XX}+3H\chi G_{3,XX}\right)\,.
\ea
The Lagrangian (\ref{Ls}) contains four nondynamical perturbations,
$\alpha$, $B$, $\delta A$, and $v$, in addition to three dynamical
fields, $\psi$, $\delta\phi$, and $\delta\rho_M$.
To derive the perturbation equations of motion, we decompose 
an arbitrary perturbation ${\cal Z}(t,{\bm x})$ in real space 
into Fourier modes ${\cal Z}_k(t)$ as
\be
{\cal Z}(t,{\bm x})=\frac{1}{(2\pi)^{3}} 
\int {\rm d}^3 k\,e^{i {\bm k} \cdot {\bm x}}
{\cal Z}_k(t)\,,
\ee
where ${\bm k}$ is the comoving wavenumber with magnitude $k = |{\bm k}|$.
In what follows, we suppress the subscript ``$k$" and work entirely 
in Fourier space. 
The equations of motion for $\alpha$, $B$, $\delta A$, 
and $v$ in Fourier space are then given, respectively, by
\ba
& &
2w_4 \alpha-w_8 \frac{\delta A}{\chi}-\frac{k^2}{a^2} 
\left( w_1 B-w_6 \psi+{\cal Y} \right)
-\dot{\phi} \dot{\delta \phi}-V_{,\phi} \delta \phi 
-\delta \rho_M=0\,,
\label{per1}\\
& &
w_1 \alpha+w_2 \frac{\delta A}{\chi}+\dot{\phi} \delta \phi
+(\rho_M+P_M)v=0\,,
\label{per2}\\ 
& &
2w_5 \delta A-w_8 \chi \alpha-\frac{k^2}{a^2} \chi
\left( w_2 B-\frac{w_2+\chi w_6}{2\chi}\psi+\frac{{\cal Y}}{2} \right)=0\,,
\label{per3}\\\
& &
\dot{\delta \rho}_M+3 \left( 1+c_M^2 \right)H \delta \rho_M
+\frac{k^2}{a^2} \left( \rho_M+P_M \right) 
\left( v+B \right)=0\,,
\label{per4}
\ea
where
\be
{\cal Y} \equiv \frac{w_3}{\chi} \left( \dot{\psi}
+\delta A+2\chi \alpha \right)\,.
\label{calY}
\ee
By solving Eqs.~(\ref{per1})-(\ref{per4}) for $\alpha$, $B$, $\delta A$, and $v$, and subsequently substituting them into Eq.~(\ref{Ss}), 
we obtain the reduced action in Fourier space in the form
\be
{\cal S}_s^{(2)}=\int \frac{{\rm d}t\,{\rm d}^3k}{(2 \pi)^3}\,a^{3}\left( 
\dot{\vec{\mathcal{X}}}^{t}{\bm K}\dot{\vec{\mathcal{X}}}
-\frac{k^2}{a^2}\vec{\mathcal{X}}^{t}{\bm G}\vec{\mathcal{X}}
-\vec{\mathcal{X}}^{t}{\bm M}\vec{\mathcal{X}}
-\vec{\mathcal{X}}^{t}{\bm B}
\dot{\vec{\mathcal{X}}} \right)\,,
\label{Ss2}
\ee
where ${\bm K}$, ${\bm G}$, ${\bm M}$, ${\bm B}$ 
are $3 \times 3$ matrices, and 
\be
\vec{\mathcal{X}}^t=\left( \psi, \delta \phi, \frac{\delta \rho_M}{k} \right)\,.
\ee
In the small-scale limit $k \to \infty$, the leading-order components of ${\bm M}$ do not contain terms proportional to $k^2$. Moreover, upon integration by parts, 
the $k^2$-dependent terms originally in ${\bm B}$ have been transferred to the corresponding components of ${\bm G}$.
The nonvanishing components of ${\bm K}$ and ${\bm G}$ 
are then given by 
\ba
K_{11} &=& 
\frac{\chi [4\Mpl^4 H^2 \chi G_{2,XX}
+\chi^3 (6\Mpl^2 H^2 + \dot{\phi}^2) G_{3,X}^2 
+ 12 \Mpl^4 H^3 (\chi^2 G_{3,XX} + G_{3,X})]}
{2(2\Mpl^2 H-\chi^3 G_{3,X})^2}\,, \nonumber \\
K_{22} &=& \frac{1}{2}\,,\qquad 
K_{12}=K_{21}=-\frac{\chi^2 \dot{\phi}\,G_{3,X}}
{2(2\Mpl^2 H-\chi^3 G_{3,X})}\,,\qquad 
K_{33}=\frac{a^2}{2(\rho_M+P_M)}\,,\\
G_{11} &=& \dot{E}_1+H E_1+\frac{\dot{\chi}}{2}G_{3,X}
+\frac{\chi^2 G_{3,X}^2 [12 \Mpl^4 H^2-3q_v \chi^2 
(\rho_M+P_M)]}
{6q_v (2\Mpl^2 H-\chi^3 G_{3,X})^2}\,, \nonumber \\
G_{22} &=& \frac{1}{2}\,,\qquad
G_{12}=G_{21}=-\frac{\chi^2 \dot{\phi}\,G_{3,X}}
{2(2\Mpl^2 H-\chi^3 G_{3,X})}\,,\qquad 
G_{33}=\frac{a^2 c_M^2}{2(\rho_M+P_M)}\,,
\ea
where 
\be
E_1 \equiv \frac{\Mpl^2 H \chi\,G_{3,X}}
{2 \Mpl^2 H-\chi^3 G_{3,X}}\,.
\ee 
The action (\ref{Ss2}) describes the propagation of three dynamical scalar perturbations, $\psi$, $\delta\phi$, and $\delta\rho_M$, which originate from the vector field, the scalar field, and the matter fluid, respectively. In the small-scale limit, the dominant contributions to Eq.~(\ref{Ss2}) come from the first two terms.

The matter perturbation $\delta \rho_M$, which is decoupled from 
the other two fields, is free of ghosts for $K_{33} > 0$, i.e., 
$\rho_M + P_M > 0$. Its squared propagation speed, given 
by $c_M^2 = G_{33}/K_{33}$, must satisfy $c_M^2 > 0$ 
to avoid Laplacian instabilities. Radiation has a pressure 
$P_r = \rho_r/3 > 0$ with $c_r^2 = 1/3$, 
so these conditions are automatically satisfied. 
For nonrelativistic matter (DM and baryons), we consider the case 
in which both $P_M$ and $c_M^2$ are positive but remain close to 0.

The ghost-free conditions for the perturbations $\psi$ 
and $\delta \phi$ are
\be
q_s \equiv K_{11}K_{22}-K_{12}^2>0\,,\quad {\rm and} \quad 
K_{22}>0\,.
\ee
Since $K_{22}=1/2$, the latter condition is automatically satisfied. 
The propagation speeds $c_s$ of the three dynamical perturbations 
can be obtained by solving ${\rm det}(c_s^2 {\bm K}-{\bm G})=0$.
One of the solutions corresponds to $c_M^2 = G_{33}/K_{33}$, 
as mentioned above.
Using the relations $G_{22} = K_{22} = 1/2$ and $G_{12} = K_{12}$, 
we find that the squared propagation speeds of $\psi$ and $\delta \phi$ 
are, respectively,
\be
c_{\psi}^2=\frac{G_{11}-2K_{12}^2}
{K_{11}-2K_{12}^2}\,,\qquad 
c_{\delta \phi}^2=1\,, 
\ee
so that $\delta \phi$ propagates at the luminal speed. 
The Laplacian instability for $\psi$ is avoided provided 
that $c_{\psi}^2 > 0$.

For the model given by the coupling functions (\ref{G23}), 
the quantities $q_s$ and $c_s^2$ can be expressed as 
\ba
\hspace{-0.8cm}
q_s &=& 2^{-\frac{s(1+p)}{1+s}} 3^{\frac{ps-1}{p(1+s)}}
\nu^{\frac{2(ps-1)}{p(1+s)}} 
\frac{m^2 p^2 s}{2(1-ps \Omega_{\chi})^2}
\left( 1+s \Omega_{\chi} \right) 
(\Omega_{\chi})^{\frac{ps-1}{p(1+s)}},
\label{qs} \\
\hspace{-0.8cm}
c_{\psi}^2 &=& \frac{6ps+5p-3+
[3-3p-2ps(2+p)]\Omega_{\chi}
-2p^2 s^2 \Omega_{\chi}^2+(2ps+p-1)
(\Omega_r+3x^2-3y^2)}{6p^2(1+s \Omega_{\chi})^2}
+\frac{2s \Omega_{\chi}}{3(1+s\Omega_{\chi})q_v u^2},
\label{cpsi}
\ea
where 
\be
q_v u^2=2^{s/(1+s)} 3^{1/[p(1+s)]} \nu_v 
\Omega_{\chi}^{1/[p(1+s)]}\,,
\label{qvu2}
\ee
with 
\be
\nu_v \equiv q_v \nu^{2/[p(1+s)]}\,.
\label{nuv}
\ee
Since we consider the case of positive energy density 
for $\chi$, i.e., $\Omega_{\chi} > 0$, 
the ghost-free condition $q_s > 0$ reduces to
\be
s>0\,.
\label{scon}
\ee
Moreover, to avoid the strong coupling problem, we require 
that $q_s$ does not approach 0 in the asymptotic past 
($\Omega_\chi \to 0$). This condition imposes the constraint 
$(ps-1)/[p(1+s)] \le 0$.
As mentioned in Eq.~(\ref{pcon}), we focus on the case $p > 0$, 
so that $u = \chi / M_{\rm pl}$ grows in time, owing to the constancy 
of $\nu=u^p H/m$.
In this case, the avoidance of strong coupling translates to 
$p s \le 1$. Under this condition, one has 
$1 - p s \Omega_{\chi} > 0$ 
for $0 < \Omega_\chi < 1$, 
ensuring that $q_s$ in Eq.~(\ref{qs}) remains finite.
Combining the condition $ps \le 1$ with (\ref{scon}), 
it then follows that 
\be
0<s \le \frac{1}{p}\,.
\label{scon1}
\ee

The last term in Eq.~(\ref{cpsi}) represents the contribution from 
an intrinsic vector mode, with $q_v = 1$ in the present model. 
Using $\Omega_\chi \propto u^{2p(s+1)}$, 
this term scales as $\Omega_\chi^{[p(s+1)-1]/[p(s+1)]}$ 
in the regime $\Omega_{\chi} \ll 1$. 
As long as
\be
p(s+1) \ge 1\,,
\label{scon2}
\ee
$c_{\psi}^2$ does not diverge in the asymptotic past. 
This condition is not strictly necessary; however, as we will show 
in Sec.~\ref{dividesec}, the positivity and finiteness of $c_\psi^2$ 
are maintained throughout the radiation, matter, and accelerated 
epochs when it is imposed together with the inequality in Eq.~(\ref{scon1}).
Imposing the condition (\ref{scon2}) in addition to (\ref{scon1}), 
the parameter $s$ is constrained to the range
\ba
\begin{cases}
0<s \le \dfrac{1}{p} &~~({\rm for}~p>1)\,, \\
\dfrac{1}{p}-1 \le s \le \dfrac{1}{p} &~~({\rm for}~0<p \le 1)\,.
\label{sp}
\end{cases}
\ea
In the following sections, we study the dynamics of the background 
and perturbations within this parameter space of $s$ and $p$.

%%%%%%%%%%%%%%%%%%%%%%%%%%%%%%%%%%%%%%
\section{Phantom-divide crossing}
\label{dividesec}
%%%%%%%%%%%%%%%%%%%%%%%%%%%%%%%%%%%%%%

Let us study the background cosmological dynamics associated with the 
coupling functions given in Eq.~(\ref{G23}). The background equations 
can be cast in autonomous form, Eqs.~(\ref{auto1})-(\ref{auto4}), 
with $h$ defined in Eq.~(\ref{h}). The fixed points relevant to the 
radiation- and matter-dominated epochs are 
\ba
& &
\text{(A)~radiation-dominated}:~(\Omega_{\chi}, x, y, \Omega_{r})=(0,0,0,1)\,,
\qquad w_{\rm eff}=\frac{1}{3}\,,\\
& &
\text{(B)~matter-dominated}:~(\Omega_{\chi}, x, y, \Omega_{r})=(0,0,0,0)\,,
\qquad w_{\rm eff}=0\,,
\ea
respectively. 
The fixed points that can give rise to late-time cosmic acceleration are
\ba
& &
\text{(C)~DE-dominated}~{\rm (i)}:~(\Omega_{\chi}, x, y, \Omega_{r})
=(1,0,0,0)\,,\qquad w_{\rm eff}=-1\,,\\
& &
\text{(D)~DE-dominated}~{\rm (ii)}:~(\Omega_{\chi}, x, y, \Omega_{r})
=\left(0,\frac{\lambda}{\sqrt{6}},\sqrt{1-\frac{\lambda^2}{6}},0 \right)\,,
\qquad w_{\rm eff}=-1+\frac{\lambda^2}{3}\,.
\ea
The point~(C) corresponds to the de Sitter solution characterized by
$h = 0$ and $w_{\rm DE} = -1$.
The point~(D), at which $h = -\lambda^2/2$ and
$w_{\rm DE} = -1 + \lambda^2/3$,
can lead to cosmic acceleration
($w_{\rm eff} = w_{\rm DE} < -1/3$) for $\lambda^2 < 2$.

To analyze the stability of the fixed points, we consider homogeneous 
perturbations $\delta \Omega_\chi$, $\delta x$, $\delta y$, and 
$\delta \Omega_r$ around them, and linearize Eqs.~(\ref{auto1})-(\ref{auto4}) as
\be
\frac{{\rm d}}{{\rm d}N} \left( \delta \Omega_{\chi}, \delta x, \delta y, 
\delta \Omega_{r} \right)^{\rm T}
={\bm A}\left( \delta \Omega_{\chi}, \delta x, \delta y, 
\delta \Omega_{r} \right)^{\rm T}\,,
\ee
where ${\bm A}$ is a $4 \times 4$ matrix. Each component of ${\bm A}$ 
is given by $A_{ij} = \partial F_i / \partial X_j$, evaluated 
at the corresponding fixed point, where $F_i$ ($i=1,2,3,4$) denote 
the right-hand sides of Eqs.~(\ref{auto1})-(\ref{auto4}), 
and $X_j = (\Omega_\chi, x, y, \Omega_r)$. 
The four eigenvalues $\mu_{1,2,3,4}$ of ${\bm A}$ 
determine the stability of each point.
For radiation- and matter-dominated fixed points, we obtain
\ba
& &
\text{(A)}~\mu_1=1\,,\quad \mu_2=2\,,\quad 
\mu_3=-1\,,\quad \mu_4=4(s+1)\,,\\
& &
\text{(B)}~\mu_1=\frac{3}{2}\,,\quad \mu_2=-\frac{3}{2}\,,
\quad \mu_3=-1\,,\quad \mu_4=3(s+1)\,,
\ea
respectively. Since we consider the case $s > 0$, 
both (A) and (B) correspond to saddle points, 
as their eigenvalues include both positive and negative values. 
For the fixed point (C), the eigenvalues are
\be
\text{(C)}~\mu_1=0\,,\quad \mu_2=-3\,,
\quad \mu_3=-3\,,\quad \mu_4=-4\,.
\ee
Although one eigenvalue vanishes at this fixed point, the system is non-hyperbolic. Expanding the equations up to second order around $(\Omega_\chi, x, y, \Omega_r) = (1,0,0,0)$, we find that the evolution of $\delta \Omega_\chi$ 
along the center direction satisfies 
$\delta \Omega_{\chi}'=-3(\delta \Omega_{\chi})^2$.
Hence, this fixed point is stable within the physical phase space. 
For the fixed point (D), the eigenvalues are
\be
\text{(D)}~\mu_1=\lambda^2-3\,,\quad \mu_2=\lambda^2-4\,,
\quad \mu_3=\frac{\lambda^2}{2}-3\,,\quad \mu_4=\lambda^2 (s+1)\,.
\ee
Under the condition for cosmic acceleration, $\lambda^2 < 2$, 
the first three eigenvalues are negative. However, $\mu_4$ is positive 
regardless of the value of $\lambda$. Therefore, the system eventually 
evolves toward the stable fixed point (C), rather than the saddle point (D). 
The cosmological evolution is thus characterized by the sequence 
of fixed points: (A) $\to$ (B) $\to$ (C).

Let us examine the evolution of $w_{\rm DE}$ from the radiation era 
to the future de Sitter epoch. As long as the DE density remains subdominant 
relative to the background fluid density, we have $h \simeq -2$ 
and $h \simeq -3/2$ during the radiation- and matter-dominated epochs, respectively.
The integrated solution to Eq.~(\ref{auto1}) during the radiation era is given by $\Omega_\chi \propto a^{4(s+1)}$, so that $\rho_\chi \propto a^{4s}$. Similarly, in the matter era, we obtain $\Omega_\chi \propto a^{3(s+1)}$, and hence $\rho_\chi \propto a^{3s}$. Since $s > 0$, the vector-field energy density $\rho_{\chi}$ 
grows with time in both regimes.

The evolution of the scalar field is governed by Eq.~(\ref{back4}). 
We consider the case in which the scalar potential $V$ contributes 
to the late-time dynamics to realize the phantom-divide crossing. 
In this case, $V$ can be estimated as $V \approx M_{\rm Pl}^2 H_0^2$, 
where $H_0$ is the present Hubble expansion rate. 
The squared mass of $\phi$ can then be estimated as
$m^2 \equiv V_{,\phi\phi} = \lambda^2 V/\Mpl^2
\approx \lambda^2 H_0^2$. For $\lambda$ at most of 
order unity, we have $m^2 \ll H^2$ during the radiation 
and matter eras ($H \gg H_0$).
Therefore, the scalar field is nearly frozen 
in these early epochs, and Eq.~(\ref{back4}) leads to
$3H\dot{\phi} \approx -V_{,\phi} = \lambda V/\Mpl$.
Using this relation, the scalar-field kinetic energy 
relative to $V$ can be estimated as
$K/V = \dot{\phi}^2/(2V) \approx \lambda^2 (H_0/H)^2$,
which implies $K \ll V$ during the early cosmological epoch.
In this regime, the scalar field energy density remains nearly  
constant, i.e., $\rho_{\phi} \simeq V \propto a^0$.

At early times, the dimensionless variable $x^2$ scales as 
$x^2 \propto H^{-4} \propto t^4$, so that $x^2 \propto a^8$ 
in the radiation era and $x^2 \propto a^6$ in the matter era. 
As long as $0 < s < 1$, $x^2$ grows faster than $\Omega_\chi$, 
and hence $\Omega_\chi$ dominates over $x^2$ in the asymptotic past. 
In the early epoch, where the conditions $x^2 \ll \Omega_\chi$ 
and $x^2 \ll y^2$ are satisfied, the DE equation of state (\ref{wde}) 
is approximately given by
\be
w_{\rm DE} \simeq -1+\frac{2hs \Omega_{\chi}}
{3(\Omega_{\chi}+y^2)}\,.
\label{wde2}
\ee
Since $\rho_\chi$ grows with time long before the onset of 
cosmic acceleration, whereas $\rho_\phi$ remains nearly constant, 
we have $\rho_\phi \gg \rho_\chi$ in the asymptotic past. 
This condition translates into $y^2 \gg \Omega_\chi$.
In this regime, the DE equation of state (\ref{wde2}) 
reduces to $w_{\rm DE} \simeq -1$. 
We denote by $z_*$ the redshift at which $\rho_\chi$ catches up 
with $\rho_\phi$, i.e., $\rho_\chi(z_*) = \rho_\phi(z_*)$. 
As long as $z_* \gg 1$, the DE equation of state in the redshift 
range $1 \ll z < z_*$ is approximately given by 
$w_{\rm DE} \simeq -1 + 2 h s/3$, which yields 
$w_{\rm DE} \simeq -1 -4s/3$ in the radiation era 
and $w_{\rm DE} \simeq -1 - s$ in the matter era.
This corresponds to the phantom regime of the DE 
equation of state ($w_{\rm DE} < -1$).
If $z_*$ is of order unity or smaller, the approximate expression
$w_{\rm DE} \simeq -1 - s$ in the matter era loses accuracy 
for $z < z_*$ due to changes in $h$, as well as the increase 
in $x^2$ induced by the onset of cosmic acceleration.

The above discussion is restricted to the early cosmological epoch, 
during which the scalar kinetic energy density is suppressed relative to 
both the vector-field energy density and 
the scalar potential ($x^2 \ll \Omega_\chi$ and $x^2 \ll y^2$). 
At later times, when the Hubble expansion rate $H$ decreases to the order
of $H_0$, the scalar field begins to evolve along its potential.
In this regime, $x^2$ can grow to be of similar order to
$\Omega_\chi$ and $y^2$. 
Under the inequality
\be
x^{2} > -\frac{h s \Omega_{\chi}}{3}\,,
\label{xcon}
\ee
the DE equation of state given by Eq.~(\ref{wde}) 
lies in the regime $w_{\rm DE}>-1$.
In particular, the phantom-divide crossing, which we denote 
by the redshift $z_c$, corresponds to the moment at which
$x^{2} = -h s \Omega_{\chi}/3$.
If the phantom-divide crossing occurs by the present time, we have $z_c > 0$. 
Whether the crossing of $w_{\rm DE} = -1$ takes place by today depends on the model parameters $s$ and $\lambda$, as well as on the initial conditions for $\Omega_\chi$, $x$, and $y$. 
Finally, the solutions evolve toward the stable de Sitter fixed
point~(C), driven by the vector-field energy density.

\begin{figure}[ht]
\begin{center}
\includegraphics[height=3.2in,width=3.4in]{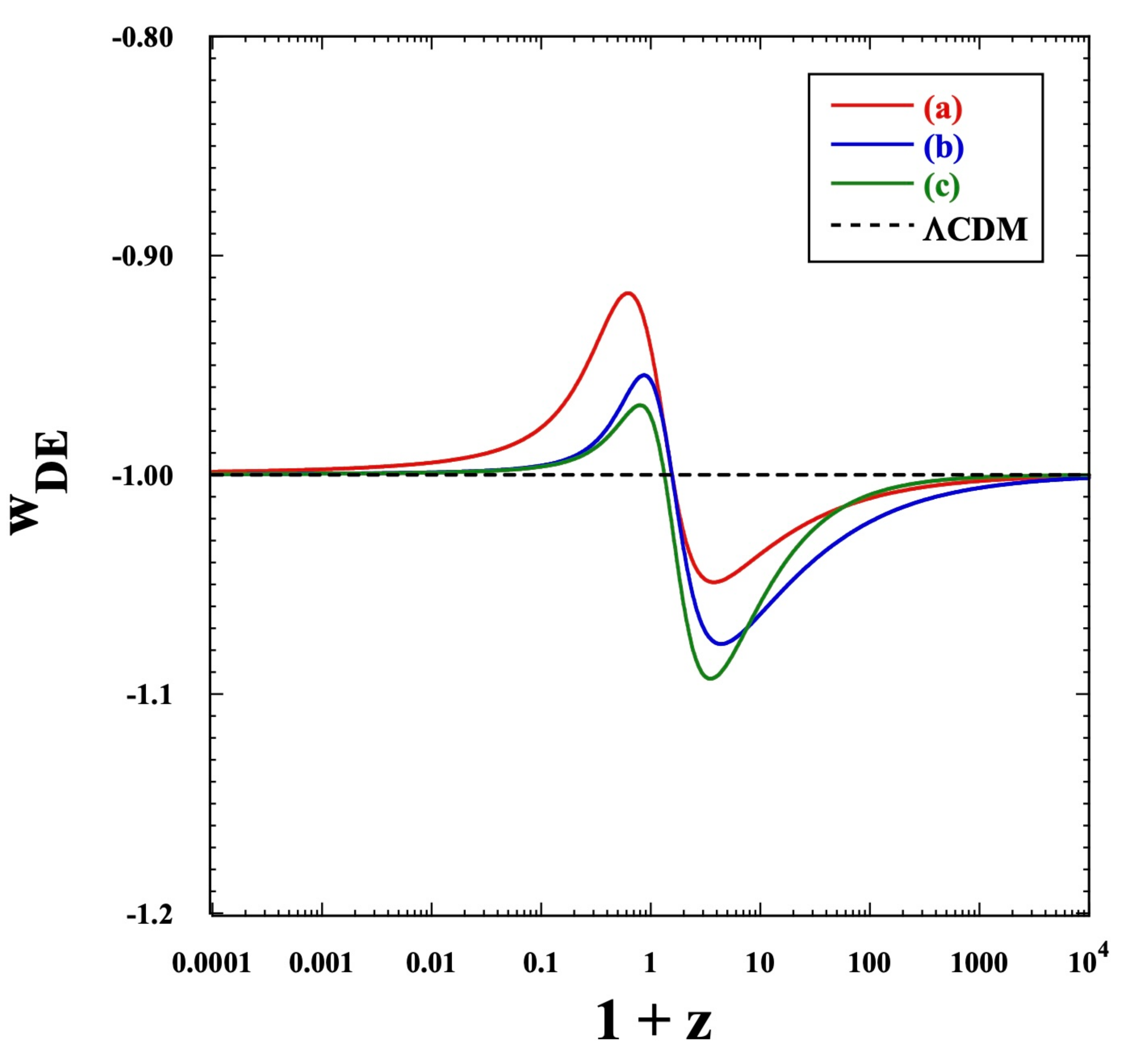}
\includegraphics[height=3.2in,width=3.4in]{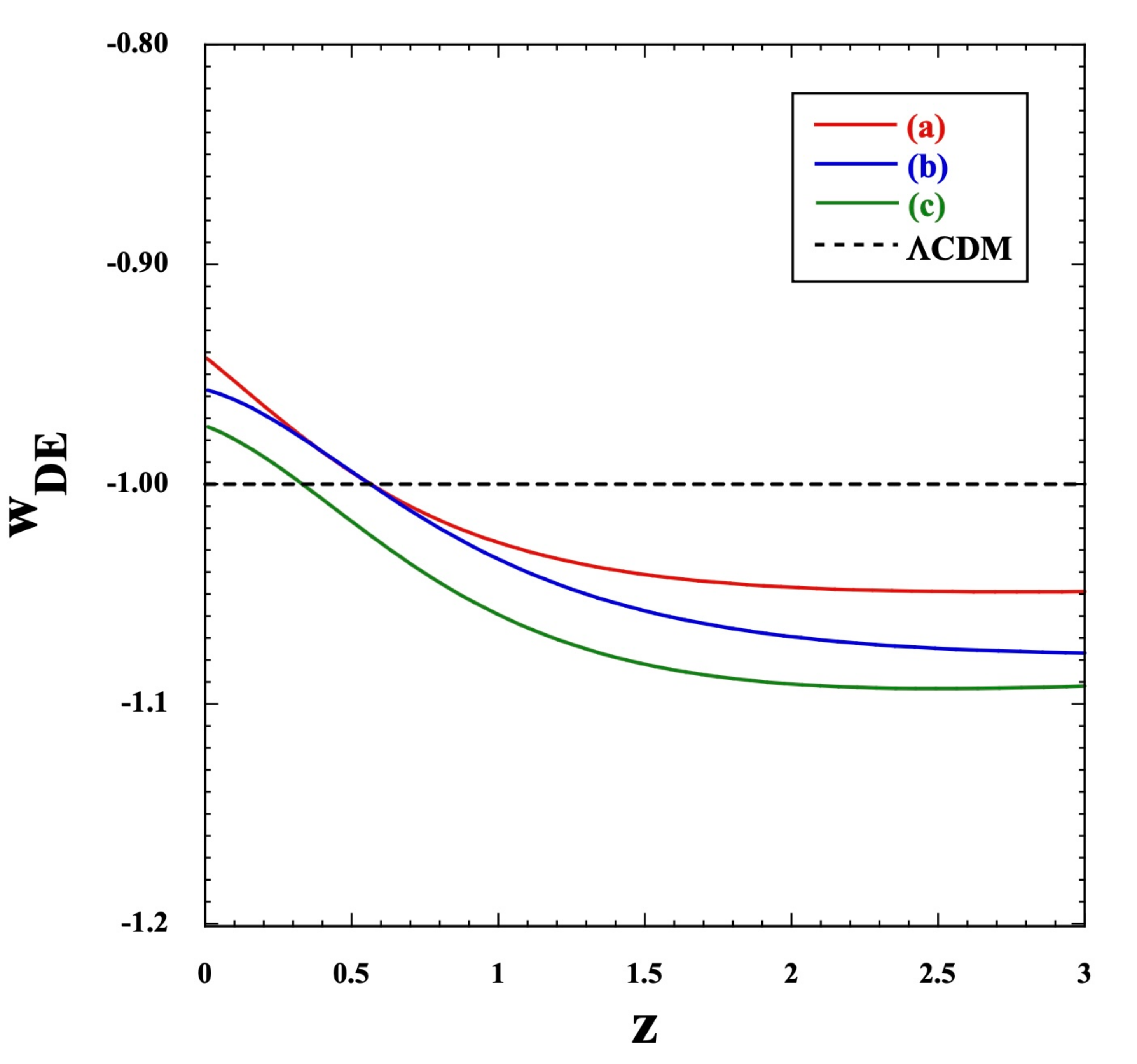}
\end{center}
\caption{Plots of $w_{\rm DE}$ as a function of the redshift $z = 1/a - 1$ are shown for the ranges $10^{-4} \le 1+z \le 10^{4}$ (left) and $0 \le z \le 3$ (right).
The scale factor is normalized such that $a=1$ today.
Each panel corresponds to the model parameters:
(a) $s = 0.2$, $\lambda = 1.5$,
(b) $s = 0.2$, $\lambda = 3$, and
(c) $s = 0.3$, $\lambda = 3$.
The initial conditions for $\Omega_{\chi}$, $x$, $y$, 
and $\Omega_r$ are given by
(a) $\Omega_{\chi} = 5.5 \times 10^{-22}$, $x = 0$, $y = 1.5 \times 10^{-9}$, $\Omega_r = 0.983957$ at $z = 2.1799 \times 10^{5}$,
(b) $\Omega_{\chi} = 2.0 \times 10^{-22}$, $x = 0$, $y = 7.0 \times 10^{-10}$, $\Omega_r = 0.988104$ at $z = 2.9524 \times 10^{5}$, and
(c) $\Omega_{\chi} = 1.1 \times 10^{-22}$, $x = 0$, $y = 2.8 \times 10^{-9}$, $\Omega_r = 0.976799$ at $z = 1.4965 \times 10^{5}$.
In all cases, the present-day values of $\Omega_{\rm DE}$ and $\Omega_r$ are $\Omega_{\rm DE} = 0.68$ and $\Omega_r = 9.0 \times 10^{-5}$, respectively.
The dashed line represents the DE equation of state in the $\Lambda$CDM model. 
\label{fig1}
}
\end{figure}

In Fig.~\ref{fig1}, we show the DE equation of state as 
a function of redshift $z$ for three cases:
(a) $s=0.2$, $\lambda=1.5$,
(b) $s=0.2$, $\lambda=3$, and
(c) $s=0.3$, $\lambda=3$. 
The initial conditions for $\Omega_{\chi}$, $x$, $y$, 
and $\Omega_r$ are given in the figure caption.
As seen in the left panel, the initial value of $w_{\rm DE}$ 
during the radiation era ($z \gtrsim 3000$) is close to $-1$, 
due to the dominance of the scalar-field energy density over 
that of the vector field ($y^2 \gg \Omega_\chi$). 
As long as $x^2$ remains subdominant relative to $\Omega_\chi$ 
and $y^2$, $w_{\rm DE}$ can be well approximated by Eq.~(\ref{wde2}). 
In the regime where $\Omega_\chi$ has not yet caught up with 
$y^2$, i.e., $\Omega_\chi \ll y^2$, Eq.~(\ref{wde2}) approximately 
reduces to $w_{\rm DE} \simeq -1 + 2hs \Omega_{\chi}/(3y^2)$,
so that $w_{\rm DE} \simeq -1 - s \Omega_{\chi}/y^2$
in the matter era. 
Since the ratio $\Omega_\chi / y^2$ increases over time,
$w_{\rm DE}$ drops below $-1$ at early epochs. 
This initial decline of $w_{\rm DE}$ is clearly seen 
in the left panel of Fig.~\ref{fig1}.

%%%%%%%%%%%%%%%%%%%%%%%%%%%%%%%%
%%%%%%%%%%%%%%%%%%%%%%%%%%%%%%%%

%%%%%%%%%%%%%%%%%%%%%%%%%%%%%%%%
\begin{figure}[ht]
\begin{center}
\includegraphics[height=3.2in,width=3.4in]{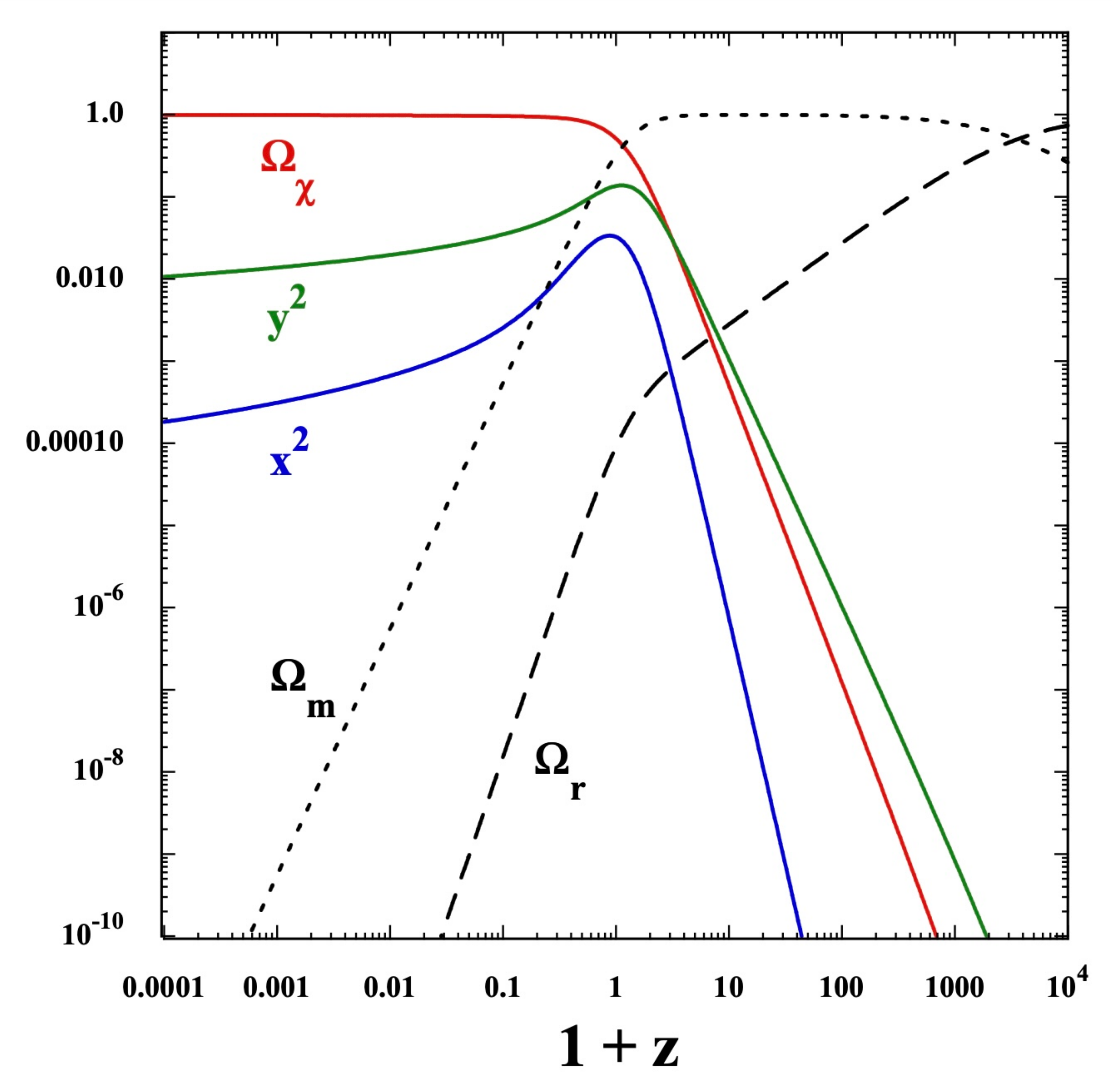}
\includegraphics[height=3.2in,width=3.4in]{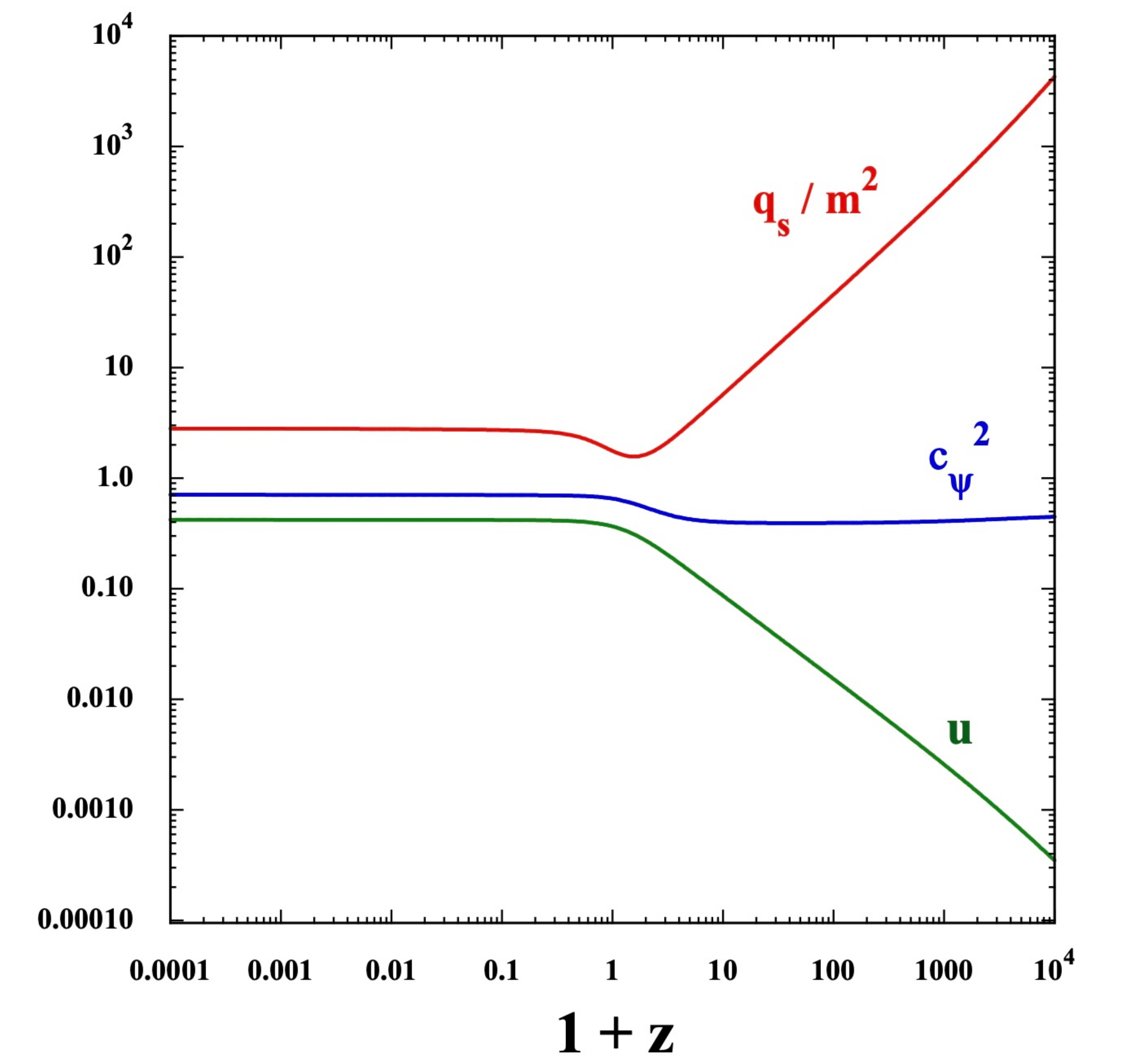}
\end{center}
\caption{Left panel: Evolution of $\Omega_\chi$, $x^2$, $y^2$, 
$\Omega_r$, and $\Omega_m$ as functions of $1+z$ 
in case (b) of Fig.~\ref{fig1}. 
Right panel: Evolution of $q_s/m^2$, $c_\psi^2$, and $u$ 
as functions of $1+z$ in case (b) of Fig.~\ref{fig1}, 
with $p=2$ and $\nu_v=\nu^{2/[p(1+s)]}=0.1$.
\label{fig2}
}
\end{figure}
%%%%%%%%%%%%%%%%%%%%%%%%%%%%%%%%

If $\Omega_{\chi}$ catches up with $y^2$ in the deep matter era, 
i.e., $z_* \gg 1$, the DE equation of state soon approaches the value 
$w_{\rm DE}=-1-s$ mentioned above. 
In such cases, however, the DE density at low redshifts 
($z<{\cal O}(1)$) is completely dominated by that of 
the vector-field energy density, i.e., $\Omega_{\chi} \gg y^2$.
Since the scalar-field kinetic energy is at most a similar order to 
$V(\phi)$, we have $\Omega_{\chi} \gg \{x^2, y^2\}$ even 
at low redshifts, and hence Eq.~(\ref{wde}) reduces to 
$w_{\rm DE} \simeq -1+2hs/3$. Since $h$ is negative 
even after the onset of cosmic acceleration, $w_{\rm DE}$ 
stays in the region $w_{\rm DE}<-1$ without the 
phantom-divide crossing. 

Therefore, to realize the phantom-divide crossing, we require
$z_* \lesssim \mathcal{O}(1)$. In case (a) of Fig.~\ref{fig1}, 
we have $z_* = -0.07$, indicating that
$\rho_\chi$ has not yet caught up with $\rho_\phi$ by today.
This implies that the DE equation of state in the early
matter-dominated epoch, which is approximately given by
$w_{\rm DE} \simeq -1 - s \Omega_\chi / y^2$ with
$\Omega_\chi \ll y^2$,
does not decrease to the value $w_{\rm DE} = -1 - s$.
Around the end of the matter-dominated era, the scalar field begins
to evolve along its potential, so that the $x^2$ term in
Eq.~(\ref{wde}) can no longer be neglected for redshifts
$z \lesssim \mathcal{O}(1)$. 
In case (a), we find that $w_{\rm DE}$ reaches a minimum at
$z = 2.8$ and starts to increase for $z < 2.8$.
In this case, the phantom-divide crossing occurs at the critical 
redshift $z_c = 0.56$ (see the right panel of Fig.~\ref{fig1}).
For $z < z_c$, the DE equation of state enters the region
$w_{\rm DE} > -1$ and subsequently reaches a maximum around
$z = -0.38$. The scalar-field kinetic energy $x^2$ plays a crucial role in
driving the evolution of $w_{\rm DE}$ from its minimum to its
maximum, with the phantom-divide crossing occurring during this
transition.
After $w_{\rm DE}$ reaches its maximum, it eventually approaches
the value $-1$, corresponding to the de Sitter attractor point~(C),
due to the dominance of $\rho_\chi$ over $\rho_\phi$. 
We note that even in the case $\lambda^2 > 2$, for which the scalar
potential alone cannot sustain cosmic acceleration, the presence
of the vector-field energy can drive accelerated expansion 
at late times. 
For instance, in case (a), the Universe enters the regime
$w_{\rm eff} < -1/3$ at redshifts $z < 0.65$. 

In case (b) of Fig.~\ref{fig1}, which corresponds to a larger value of
$\lambda$ than in case (a) but with the same value of $s$, the minimum
value of $w_{\rm DE}$ reached around $z = 3.4$ is smaller than that in
case (a). This reflects the fact that, in case (b), the ratio
$\Omega_\chi / y^2$ at the same redshift is larger during the early
matter-dominated era, leading to a smaller value of
$w_{\rm DE} \simeq -1 - s \Omega_\chi / y^2$. 
In this case, we have $z_*=2.0$, so that $\rho_{\chi}$ catches up 
with $\rho_{\phi}$ in the past. 
This behavior can be seen in the left panel of Fig.~\ref{fig2},
where $\Omega_\chi$ grows to be larger than $y^2$ around
$z = 2$. Since $\Omega_\chi$ is smaller than $y^2$ at $z = 3.4$, the minimal
value of $w_{\rm DE}$ is larger than $w_{\rm DE} = -1 - s = -1.2$,
namely $w_{\rm DE,min} = -1.08$.
After $w_{\rm DE}$ reaches its minimum, the growth of the scalar-field
kinetic energy $x^2$ leads to an increase in $w_{\rm DE}$, which is
followed by the phantom-divide crossing at $z_c = 0.56$.
Even though $x^2$ is smaller than $\Omega_\chi$ and $y^2$,
the condition~(\ref{xcon}) can be satisfied for $s$ of order $0.1$.
In Fig.~\ref{fig1}, we find that the maximum value of $w_{\rm DE}$ in
case (b) is smaller than that in case (a).
Since the slope parameter $\lambda$ in case~(b) is larger,
the potential energy $V(\phi)$ starts to decrease more rapidly.
After $w_{\rm DE}$ reaches its maximum, the Universe quickly approaches
the de Sitter fixed point dominated by the vector-field energy density.

In case (c) of Fig.~\ref{fig1}, the slope parameter $\lambda$ is the
same as in case (b), but the value of $s$ is larger. 
The minimum value of $w_{\rm DE}$, which is realized at $z = 2.5$, is
smaller than that in case (b), reflecting the fact that
$w_{\rm DE}$ in Eq.~(\ref{wde2}) tends to deviate further from $-1$
as $s$ increases. In case (c), the phantom-divide crossing occurs 
at $z_c=0.33$, with $z_*=1.30$. 
In the right panel of Fig.~\ref{fig1}, we find that the deviation of
$w_{\rm DE}$ from $-1$ prior to the phantom-divide crossing is
largest in case (c), compared with cases (a) and (b). 
After $w_{\rm DE}$ crosses $-1$, however, the maximum value reached
in the regime $w_{\rm DE} > -1$ is smallest in case (c). 
This behavior can be attributed to the fact that, for larger values of $s$ and
$\lambda$, the vector-field energy density $\rho_\chi$ tends to
dominate more strongly over the scalar-field energy density 
$\rho_\phi$ at late times.

If we further increase the value of $s$, for example to $s > 0.5$,
the condition (\ref{xcon}) required for the phantom-divide crossing
becomes more difficult to satisfy.
Since the dominance of $\Omega_\chi$ over $y^2$ should not persist
down to low redshifts for the crossing of $w_{\rm DE} = -1$ to occur, 
the DE equation of state $w_{\rm DE} \simeq -1 - s \Omega_\chi / y^2$ 
in the early matter-dominated era does not reach the value
$w_{\rm DE} \simeq -1 - s$.
Requiring that the phantom-divide crossing occurs by today,
we numerically find that $w_{\rm DE}$ does not strongly
deviate from $-1$ (e.g., it does not typically enter a region
such as $w_{\rm DE} < -1.5$) before crossing $w_{\rm DE} = -1$.
For the parameters $s = \mathcal{O}(0.1)$ and
$\lambda = \mathcal{O}(1)$, cases (a), (b), and (c) shown in
Fig.~\ref{fig1} provide typical examples of the evolution of
$w_{\rm DE}$. By choosing values of $s$ and $\lambda$ closer to $0$, 
the evolution of $w_{\rm DE}$ approaches that of the $\Lambda$CDM model. 

For cases in which the phantom-divide crossing occurs, the DE equation of state exhibits a minimum in the past. Such an evolution of $w_{\rm DE}$ cannot be captured by the CPL parameterization, $w_{\rm DE} = w_0 + w_a z/(1+z)$, 
as it can describe only a monotonically increasing or decreasing behavior of $w_{\rm DE}(z)$. 
Therefore, confronting our model with observations requires solving the
autonomous Eqs.~\eqref{auto1}-\eqref{auto4} to determine
$w_{\rm DE}(z)$, rather than relying on the CPL parameterization.

We also examine whether the linear stability
conditions derived in Sec.~\ref{linearsec} are satisfied for the
background evolution shown in Fig.~\ref{fig1}.
In the right panel of Fig.~\ref{fig2}, we plot the evolution of
$q_s/m^2$, $c_\psi^2$, and $u$ for case (b) of Fig.~\ref{fig1},
with $p = 2$ and $\nu_v = \nu^{2/[p(1+s)]} = 0.1$.
We find that $q_s$ is always positive and increases toward the
asymptotic past, corresponding to the weak-coupling
limit. In the future, $q_s$ approaches a finite positive constant. 
Recalling that the phantom-divide crossing occurs at
$\Omega_\chi = -3x^2/(h s) > 0$, we see from Eq.~(\ref{qs}) that
$q_s$ remains positive and finite across the crossing 
of $w_{\rm DE} = -1$. As confirmed in Fig.~\ref{fig2}, 
the ghost-free condition is satisfied from the radiation era 
to the future de Sitter attractor, without any discontinuities 
of $q_s$ and $u$. The temporal vector component $u$ grows 
with time in the region $u>0$, so that $\Omega_{\chi}>0$ 
in Eq.~(\ref{Omechi}).

In the right panel of Fig.~\ref{fig2}, we also find that $c_\psi^2$ is
positive throughout the cosmological evolution starting from the
radiation era. From Eq.~\eqref{cpsi}, the values of $c_\psi^2$ 
at the fixed points (A), (B), and (C) are given, respectively, by 
\ba
c_{\psi}^2=
\begin{cases}
\dfrac{(4s+3)p-2}{3p^2} &~~[\text{radiation point (A)}]\,, \\
\dfrac{(6s+5)p-3}{6p^2} &~~[\text{matter point (B)}]\,, \\
\dfrac{1-ps}{3p(s+1)}+\dfrac{1}{3\nu_v} 
\left( \dfrac{2}{3^{1/p}} \right)^{1/(1+s)} \dfrac{s}{1+s} 
&~~[\text{de Sitter point (C)}]\,.
\label{cpsi2}
\end{cases}
\ea
Under the condition~\eqref{sp} with $\nu_v > 0$, the three values of
$c_\psi^2$ in Eq.~\eqref{cpsi2} are all positive.
The numerical simulation in Fig.~\ref{fig2} indicates that $c_\psi^2$ also remains positive during the transient epochs between the three periods. Since we have numerically confirmed that $c_\psi^2$ and $q_s$ are positive in cases (a) and (c) of Fig.~\ref{fig1} as well, there are neither ghost nor Laplacian instabilities for the longitudinal scalar perturbation $\psi$ arising from the breaking of the $U(1)$ gauge symmetry. 

At the end of this section, we point out the dependence of the phantom-divide crossing behavior on the choice of initial conditions for $\Omega_{\chi}$, $x$, and $y$ at high redshifts.
As the initial value of $\Omega_{\chi}$ increases, the critical redshift $z_c$ at which $w_{\rm DE} = -1$ is reached tends to decrease, due to the dominance of the vector-field energy density at low redshifts.
For sufficiently large initial $\Omega_{\chi}$, the crossing of $w_{\rm DE} = -1$ may not occur by the present time.
The dependence of $z_c$ on the initial value of $x$ is fairly weak, as long as $x$ is much smaller than unity. This is because the scalar-field kinetic energy initially decreases due to Hubble friction as the field evolves along the potential.
For larger initial values of $y$, $z_c$ tends to increase due to the earlier dominance of the scalar-field potential.
For given $s$ and $\lambda$ in the ranges $0 < s < 0.5$ and 
$\lambda = \mathcal{O}(1)$, there typically exist initial conditions for $\Omega_{\chi}$, $x$, and $y$ under which the phantom-divide crossing occurs at $0 < z_c < 1$.
In such cases, the present-day values of $\Omega_{\chi}$, $x$, and $y$ are typically of order $0.1$; for example, $\Omega_{\chi}(z=0) = 0.5119$, $x(z=0) = 0.1804$, and $y(z=0) = 0.3683$ in case (b) of Fig.~\ref{fig1}.
We leave a detailed analysis of observational constraints on the model, based on the background evolution of $w_{\rm DE}$, for future work.

%%%%%%%%%%%%%%%%%%%%%%%%%%%%%%%%%%%%%%
\section{Growth of inhomogenities}
\label{growthsec}
%%%%%%%%%%%%%%%%%%%%%%%%%%%%%%%%%%%%%%

Let us study the evolution of quantities associated with observables
in measurements of the growth of inhomogeneities. 
For this purpose, we consider perturbations of nonrelativistic matter
(CDM and baryons), which are characterized as pressureless matter
($P_m = 0$) with vanishing sound speed ($c_m^2 = 0$). 
Since we are interested in the dynamics of perturbations
in the late Universe, we neglect the contribution of radiation
to both the background and the perturbations.
From the matter density perturbation $\delta \rho_m$ and
the velocity potential $v$, we define the gauge-invariant
density contrast $\delta_m$ as
\be
\delta_m \equiv \frac{\delta \rho_m}{\rho_m} + 3 H v \, .
\label{deltam}
\ee
Varying the second-order Lagrangian~(\ref{Ls}) for scalar perturbations
with respect to $\delta \rho_m$, where the subscript $M$ is replaced
by $m$, we obtain
\be
\dot{v}=\alpha\,.
\label{per5}
\ee
We also find that Eq.~(\ref{per4}) reduces to 
\be
\dot{\delta}_m+\frac{k^2}{a^2} \left( v+B \right) 
-3\dot{\cal V}=0\,,
\label{delmeq0}
\ee
where ${\cal V} \equiv Hv$. 

We also introduce the gauge-invariant gravitational potentials as
\be
\Psi \equiv \alpha+\dot{B}\,,\qquad 
\Phi \equiv -H B\,.
\ee
Differentiating Eq.~\eqref{delmeq0} with respect to $t$
and using Eq.~\eqref{per5}, we obtain
\be
\ddot{\delta}_m+2H \dot{\delta}_m+\frac{k^2}{a^2}\Psi
=3 \left( \ddot{\cal V} +2H \dot{{\cal V}} \right)\,.
\label{delmeq}
\ee
We define the dimensionless quantity $\mu$ through
the modified Poisson equation,
\be
\frac{k^2}{a^2} \Psi=-4 \pi G_{\rm N} \mu\,
\rho_m \delta_m\,,
\label{Psia}
\ee
where $G_{\rm N} = (8\pi M_{\rm pl}^2)^{-1}$ is the 
Newton gravitational constant. 
While $\mu = 1$ in General Relativity (GR), deviations from 
1 generally arise in theories beyond GR.
Such a modification affects the gravitational clustering
of matter density perturbations through Eq.~\eqref{delmeq}.
Measurements of matter growth rates, such as those from redshift-space
distortions \cite{Blake:2011rj,Beutler:2012px,delaTorre:2013rpa,Howlett:2014opa,Okumura:2015lvp}, allow one to place constraints on $\mu$.

The combination of gravitational potentials relevant for observations
in weak lensing and ISW-galaxy cross-correlations is 
given by 
$\Psi+\Phi$ \cite{Amendola:2007rr,Bertschinger:2008zb,Tsujikawa:2008in,Zhao:2008bn,Zhao:2010dz,Song:2010fg,Kimura:2011td,Simpson:2012ra,Bolis:2018vzs,Nakamura:2018oyy,Kable:2021yws}.
This effect can be quantified by introducing the dimensionless
parameter $\Sigma$, defined through
\be
\frac{k^2}{a^2}\left(\Psi + \Phi\right)
= -8\pi G_{\rm N} \Sigma\,
\rho_m \delta_m \,.
\label{Psia2}
\ee
The deviation of $\Sigma$ from 1 characterizes the modification
of light-ray bending relative to the GR case.

Varying the second-order Lagrangian (\ref{Ls}) of scalar perturbations 
with respect to $\psi$ and $\delta \phi$, we obtain
\ba
& &
\dot{\cal Y}+\left( H -\frac{\dot{\chi}}{\chi} \right) {\cal Y}
+\frac{1}{\chi} \left[ \chi^2 (2w_6 \alpha+2w_7 \psi)
+(w_2+w_6 \chi) \delta A \right]=0\,,\label{per6}\\
& &
\ddot{\delta \phi}+3H \dot{\delta \phi}
+\left( \frac{k^2}{a^2}+V_{,\phi \phi} \right) 
\delta \phi+2V_{,\phi}\alpha-\dot{\phi} \dot{\alpha}
+\frac{k^2}{a^2}\dot{\phi}B=0\,,\label{per7}
\ea
where ${\cal Y}$ is defined by Eq.~(\ref{calY}).
By numerically solving the perturbation equations of motion
(\ref{per1})-(\ref{per4}), (\ref{per5}), (\ref{per6}), and (\ref{per7})
for a given wavenumber $k$, we obtain the time evolution of
$\mu$ and $\Sigma$.

\subsection{Quasi-static approximation}

In observations of redshift-space distortions, weak lensing, and
ISW-galaxy cross-correlations, we are primarily interested in how
$\mu$ and $\Sigma$ evolve in time at low redshifts for perturbations
deep inside the Hubble radius, $k^2 \gg a^2 H^2$. 
To this end, we resort to the so-called quasi-static approximation,
under which the dominant contributions to the perturbation equations
are those containing $k^2/a^2$ and $\delta\rho_m$
\cite{Boisseau:2000pr,Tsujikawa:2007gd,Nesseris:2008mq,Song:2010rm,DeFelice:2011hq}.
Strictly speaking, this approximation is valid for modes deep inside
the sound horizon of DE perturbations.
In our model, the scalar field has a luminal sound speed, whereas the
squared propagation speed of the longitudinal mode of the vector field
is given by $c_\psi^2$.
As shown in the right panel of Fig.~\ref{fig2}, we mainly focus on the
case in which $c_\psi^2$ does not significantly deviate from 1.
In such cases, the DE sound horizon is of the same order as 
the Hubble radius.

Under the quasi-static approximation, the perturbation 
Eqs.~(\ref{per1}), (\ref{per3}), and (\ref{per7}) reduce, respectively, to
\ba
& & \frac{k^2}{a^2} \left( \frac{w_1}{H} \Phi 
+w_6 \psi-{\cal Y}  \right)-\delta \rho_m=0\,,\label{per1a}\\
& &  -\frac{w_2}{H}\Phi-\frac{w_2+\chi w_6}{2\chi}\psi
+\frac{{\cal Y}}{2}=0\,,\label{per3a}\\
& & \delta \phi-\frac{\dot{\phi}}{H}\Phi=0\,,
\label{per7a}
\ea
where we have replaced $B$ with $-\Phi/H$.
To obtain Eq.~(\ref{per7a}), we have neglected the mass-squared term
$m_s^2 = V_{,\phi\phi}$ of the scalar field $\phi$.
Since we consider the case in which $m_s$ is of the same order as $H_0$,
this approximation is justified for modes deep inside 
today's Hubble radius, i.e., $k^2/a^2 \gg H_0^2 \approx m_s^2$.
Eliminating ${\cal Y}$ from Eqs.~(\ref{per1a}) and (\ref{per3a}),
we obtain
\be
\delta \rho_m=\frac{k^2}{a^2} \left( \frac{w_1-2w_2}{H} \Phi
-\frac{w_2}{\chi} \psi \right) \,.
\label{drhom}
\ee
Substituting the definition of ${\cal Y}$ in Eq.~(\ref{calY})
into Eq.~(\ref{per3a}) yields
\be
\dot{\psi}=-2 \chi \alpha -\delta A+\frac{1}{w_3} 
\left[ (w_2+\chi w_6) \psi+\frac{2\chi w_2}{H}\Phi \right]\,.
\label{dotpsi}
\ee
From Eqs.~(\ref{per2}) and (\ref{per4}), the velocity potential $v$
and the nonrelativistic matter density perturbation $\delta\rho_m$
respectively obey
\ba
& &
v=-\frac{1}{\rho_m} \left[ w_1 \left( \Psi+\frac{\dot{\Phi}}{H}
-\frac{\dot{H}}{H^2}\Phi \right)+\frac{w_2}{\chi} \delta A 
+\dot{\phi} \delta \phi \right]\,,\label{per2a}\\
& &
\dot{\delta \rho}_m+3H \delta \rho_m
+\frac{k^2}{a^2} \rho_m \left( v-\frac{\Phi}{H} \right)=0\,.
\label{per4a}
\ea
Substituting Eq.~(\ref{drhom}) and its time derivative, together with
Eq.~(\ref{per2a}), into Eq.~(\ref{per4a}), and exploiting 
Eqs.~(\ref{per7a}) and (\ref{dotpsi}) to eliminate $\delta \phi$ 
and $\dot{\psi}$, we find\footnote{The parameters $\kappa_{1,2,3,4,5}$ used in 
Ref.~\cite{Kase:2018nwt} correspond to $\kappa_1=\beta_1$, 
$\kappa_2=\beta_2+\dot{\phi}^2/H$, 
$\kappa_3=\beta_3$, $\kappa_4=0$, and $\kappa_5=\beta_4$ 
in our notations of $\beta_{1,2,3,4}$, with the replacements
$\Phi\to -\Phi$ and $A_0\to -\chi$.}
\be
\beta_1 \Psi-\beta_2 \Phi+\frac{\beta_3}{w_3 \chi^2} 
\psi=0\,,
\label{quasi1}
\ee
where 
\be
\beta_1 \equiv w_1-2w_2\,,\qquad 
\beta_2 \equiv \frac{1}{H} \left( \dot{\beta}_1
+H \beta_1-\rho_m-\frac{2w_2^2}{w_3}
-\dot{\phi}^2 \right)\,,\qquad 
\beta_3 \equiv w_2 w_6 \chi^2
+\left( \dot{w}_2 w_3 +H w_2 w_3+w_2^2 \right)\chi
-w_2 w_3 \dot{\chi}\,.
\ee
We also solve Eq.~(\ref{per3a}) for $\mathcal{Y}$ and take its time derivative.
Substituting these expressions into Eq.~(\ref{per6}) and using Eq.~(\ref{dotpsi}),
we obtain
\be
2 w_2 w_3 \chi^2 \Psi - \frac{2 \chi}{H} \beta_3 \Phi
+ \beta_4 \psi=0\,,
\label{quasi2}
\ee
where 
\be
\beta_4 \equiv -(2w_3 w_7+w_6^2) \chi^3
-\left[ (\dot{w}_6+H w_6)w_3+2w_2 w_6 \right] \chi^2
-\left[ (\dot{w}_2+H w_2-w_6 \dot{\chi})w_3+w_2^2 \right] \chi
+2 w_2 w_3 \dot{\chi}\,.
\ee
Solving Eqs.~(\ref{drhom}), (\ref{quasi1}), and (\ref{quasi2}) 
for $\Psi$, $\Phi$, and $\psi$, it follows that 
\ba
\Psi &=& \frac{H \beta_2 \beta_4 \chi w_3 
- 2 \beta_3^2}{2\Delta} \frac{a^2}{k^2} \delta \rho_m\,,\label{Phiap}\\
\Phi &=& \frac{H (\beta_1 \beta_4 - 2 \beta_3 w_2) w_3 \chi}
{2\Delta} \frac{a^2}{k^2} \delta \rho_m\,,\label{Psiap}\\
\psi &=& \frac{(\beta_1 \beta_3-H \chi \beta_2 w_2 w_3)w_3 \chi^2}
{\Delta}\frac{a^2}{k^2} \delta \rho_m\,,
\ea
where
\be
\Delta \equiv \frac{1}{2} w_3 \chi \left( 
\beta_1^2 \beta_4-4\beta_1 \beta_3 w_2 
+2H \chi \beta_2 w_2^2 w_3 \right)\,.
\ee
Using the properties $w_3=-2\chi^2 q_v$, $w_6=w_2/\chi$, 
$w_7=\dot{\chi}w_2/\chi^3$, and the background equation 
of motion, we obtain the following relation:
\be
q_v H \chi^3 \left[ 2\beta_3 w_2-\beta_4 (\beta_1-\beta_2) 
\right]+\beta_3^2=0\,.
\label{muSire}
\ee
Comparing Eqs.~(\ref{Phiap}) and (\ref{Psiap}) with
Eqs.~(\ref{Psia}) and (\ref{Psia2}), and using the relation
(\ref{muSire}), we find that $\mu$ and $\Sigma$ coincide 
and are given by
\be
\mu=\Sigma=\frac{H \chi^3 \beta_2 \beta_4 q_v+\beta_3^2}
{4\pi G_{\rm N} \Delta}\,,
\ee
where $\Delta$ is equilavent to 
\be
\Delta=16 \beta_1^2 \chi^8 q_v^2 q_s c_{\psi}^2\,.
\ee
In particular, for the coupling functions given in
Eq.~(\ref{G23}), we obtain the following concise formula:
\be
\mu=\Sigma=1+\frac{s \Omega_{\chi}}
{3(1+s\Omega_{\chi})c_{\psi}^2}\,.
\label{muana}
\ee
Since we require $s>0$ and $c_{\psi}^2>0$ to avoid ghosts and
Laplacian instabilities, it follows that $\mu=\Sigma>1$.

\subsection{Growth of matter perturbations}

Let us study the evolution of $\mu$ using the expression 
given in Eq.~(\ref{muana}). For modes well inside the Hubble 
radius, the terms on the right-hand side of Eq.~(\ref{delmeq}) 
are negligible compared with those on the left-hand side.
In this regime, the matter perturbation approximately satisfies
\be
\ddot{\delta}_m
+ 2H \dot{\delta}_m
- 4\pi G_{\rm N} \mu \rho_m \delta_m = 0 \,.
\ee
Since $\mu>1$, the growth rate of the matter 
density contrast
is enhanced relative to that in the $\Lambda$CDM model. 
In the early Universe, we have $\Omega_\chi \ll 1$ in
Eq.~(\ref{muana}), so that $\mu$ is close to 1. 
As $\Omega_\chi$ grows to ${\cal O}(0.1)$ 
at low redshifts, $\mu$ begins to deviate from 1.
For smaller values of $c_\psi^2$, $\mu$ tends to be larger.
From Eq.~(\ref{cpsi}), we find that $c_\psi^2$ is affected by
the transverse vector mode through the quantity 
$\nu_v=q_v \nu^{2/[p(1+s)]}$, 
where $q_v = 1$ in our model, and
$\nu = (\chi/M_{\rm pl})^p (H/m)$ is constant due to the relation (\ref{chipre}).

We recall that the background autonomous Eqs.~(\ref{auto1})-(\ref{auto4}) depend only on the parameter $s = p_2/p$ and the parameter $\lambda$ of the exponential potential. Since there is a freedom to choose the value of $p$ independently of $s$ and $\lambda$, the parameter $\nu_v$ can also be specified independently of the background.
For larger $\nu_v$, $c_\psi^2$ becomes smaller, leading to higher values of $\mu$. The limit $\nu_v \to \infty$ corresponds to a weakly coupled regime between the longitudinal and transverse vector modes, so that the propagation of the longitudinal mode mimics that of scalar-field perturbations in scalar-tensor theories \cite{DeFelice:2010as,Kimura:2011td,Barreira:2012kk}.
For smaller $\nu_v$, the longitudinal vector perturbation tends to be more strongly affected by the transverse mode. As a result, deviations from scalar-tensor theories become manifest, with $\mu$ and $\Sigma$ approaching 1 in the limit $\nu_v \to 0$.

\begin{figure}[ht]
\begin{center}
\includegraphics[height=3.2in,width=3.4in]{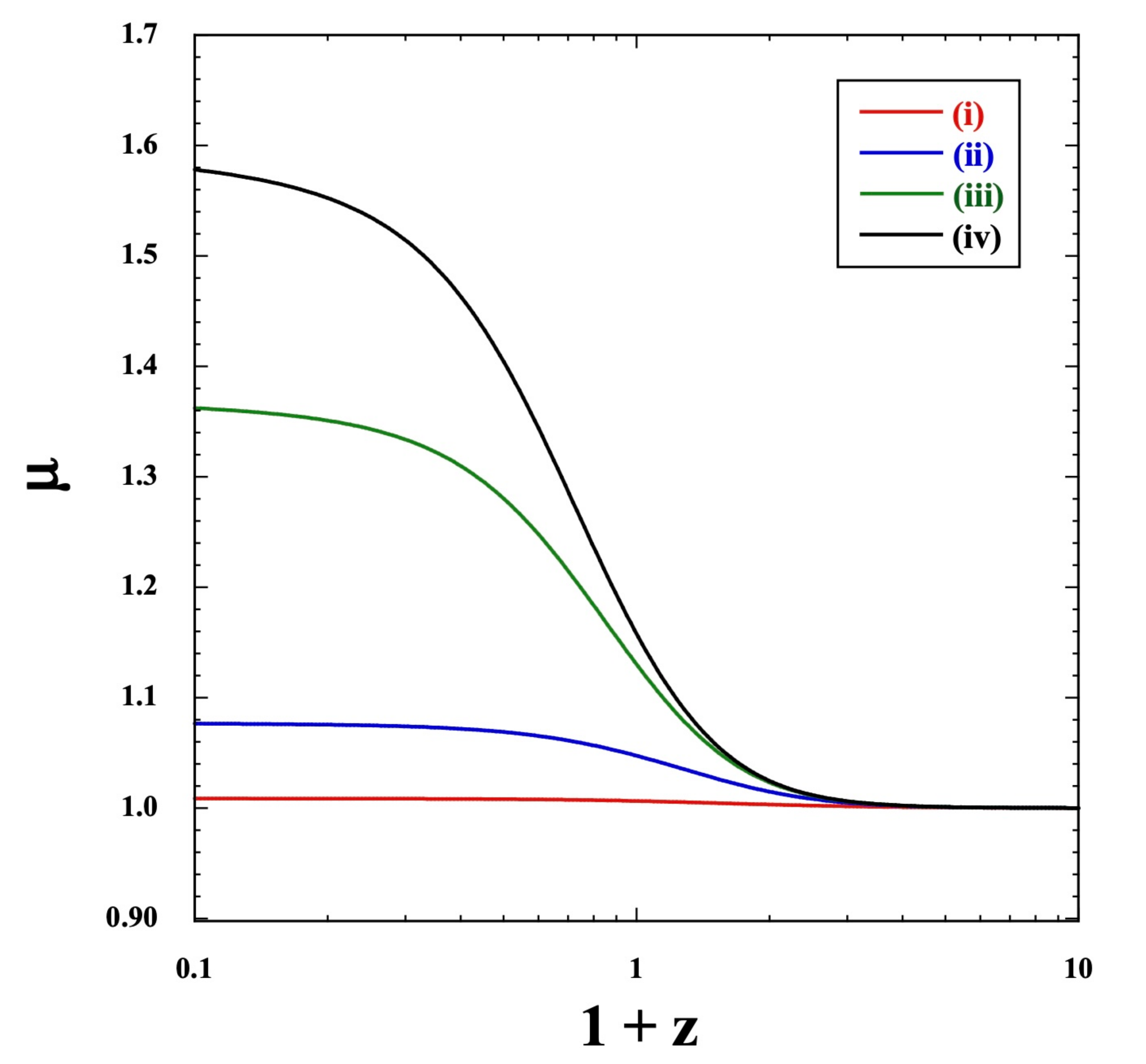}
\includegraphics[height=3.2in,width=3.4in]{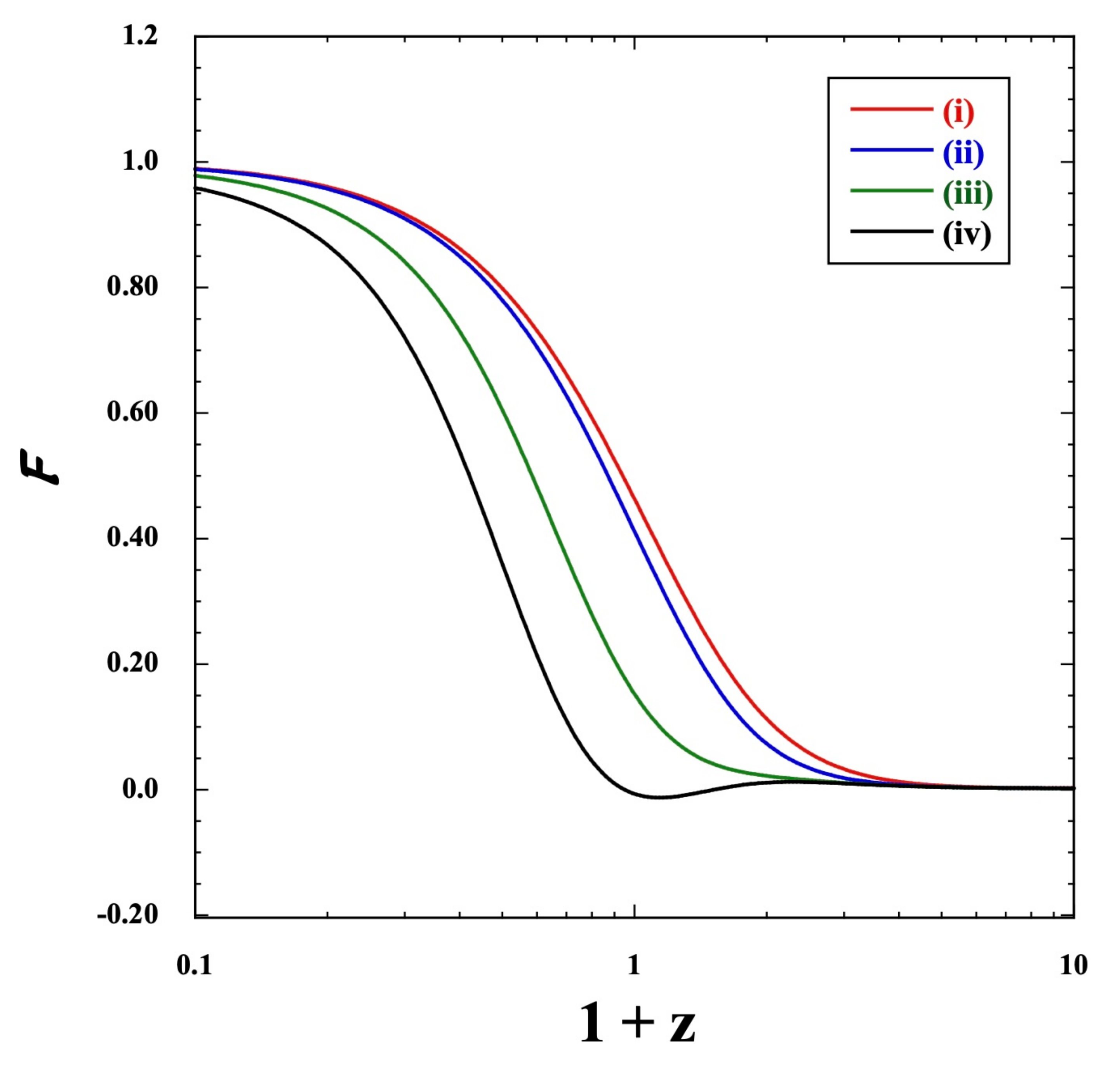}
\end{center}
\caption{Left panel: Evolution of the quantity $\mu$ as a function
of $1+z$ for case~(b) of Fig.~\ref{fig1}, with $p=2$.  
Each line corresponds to (i) $\nu_v=0.01$, (ii) $\nu_v=0.1$, 
(iii) $\nu_v=1$, and (iv) $\nu_v=10$.  
Right panel: Evolution of the quantity ${\cal F}$ as a function
of $1+z$ for case~(c) of Fig.~\ref{fig1}, with $p=2$.  
Each line corresponds to (i) $\nu_v=0.01$, (ii) $\nu_v=0.1$, 
(iii) $\nu_v=1$, and (iv) $\nu_v=10$.
\label{fig3}}
\end{figure}

In the left panel of Fig.~\ref{fig3}, we plot the evolution of $\mu$
for the background corresponding to case~(b) in Fig.~\ref{fig1},
considering four different values of $\nu_v$. 
For $\nu_v = 0.01$, the deviation of $\mu$ from 1 at low redshifts 
is small, i.e., $\mu - 1 \ll 1$. In this case, the growth of $\delta_m$ 
is similar to that in the $\Lambda$CDM model, apart from 
differences arising from modifications to the background. 
For $\nu_v \gtrsim \mathcal{O}(0.1)$, deviations of $\mu$ from 
the $\Lambda$CDM model begin to appear at low redshifts. 
The present-day values of $\mu$ for $\nu_v = 0.1, 1$, 
and $10$ are $\mu = 1.05, 1.13$, and $1.16$, respectively.  
Since a significantly larger enhancement of $\delta_m$ than 
in the $\Lambda$CDM model can be constrained by growth-rate 
measurements such as redshift-space distortions
\cite{Blake:2011rj,Beutler:2012px,delaTorre:2013rpa,Howlett:2014opa,Okumura:2015lvp}, it should be possible to place upper bounds on the values of $\nu_v$.  
As long as $\nu_v \ll 1$, the cosmic growth history closely mimics that
in the $\Lambda$CDM model.

The advantage of considering a vector field with a broken $U(1)$ gauge symmetry,
compared to scalar-tensor theories, is that the presence of transverse
vector modes affects the propagation speed $c_\psi$ of the longitudinal
scalar perturbation. 
As discussed above, this allows $\mu$ to take values closer 
to 1 under the condition $\nu_v \ll 1$.
Such a property does not hold for scalar-tensor theories with
Galileon-type self-interactions, in which the deviation of $\mu$
from 1 at low redshifts is generally significant \cite{DeFelice:2010as,Barreira:2012kk,Renk:2017rzu}.
If we break the shift symmetry of scalar Galileons by introducing
a scalar potential $V(\phi)$, the large growth rate of $\delta_m$
can be reduced by suppressing the dominance of the Galileon energy
density at low redshifts \cite{Tsujikawa:2025wca}.
Indeed, a similar suppression of $\mu$ occurs in our scalar-vector-tensor
theories, as the presence of the potential $V(\phi)$ leads to smaller
values of $\Omega_\chi$ in Eq.~(\ref{muana}) compared to GP theories.
Numerically, we have confirmed that the values of $\mu$ are generally
smaller than those in GP theories with $V(\phi)=0$ by choosing the
same model parameters.
Thus, the presence of the scalar potential in our model, as well as the intrinsic
vector mode with $\nu_v \ll 1$, allows a flexible possibility for realizing values
of $\mu$ and $\Sigma$ close to 1. 

%%%%%%%%%%%%%%%%%%%%%%%%%%%%%%%%
%%%%%%%%%%%%%%%%%%%%%%%%%%%%%%%%

%
\subsection{Compatibility with ISW-galaxy cross-correlations}

The ISW-galaxy cross-correlations \cite{Boughn:2003yz,Afshordi:2003xu,Corasaniti:2005pq,Pogosian:2005ez}
can also probe signatures of large-scale modifications of gravity.
Since the ISW effect in the CMB is directly related to the bending of light rays,
the quantity $\Sigma$ in Eq.~(\ref{muana}) affects the cross-correlation spectrum
between the ISW signal and galaxy clustering.
To characterize these ISW-galaxy cross-correlations, 
we introduce the growth factor
$D(z)$ at redshift $z$, defined by
\be
\delta_m (z,{\bm k})=\frac{D(z)}{D_0} \delta_m (0,{\bm k})\,,
\ee
where $D_0$ denotes the present-day value of $D$.
We consider galaxy catalogs with a window function $W$ 
and a bias factor $b_s$, and introduce a comoving distance 
$d_c(z)=\int_0^{z} H^{-1} (\tilde{z}){\rm d} \tilde{z}$.
Using the so-called Limber approximation, the 
ISW-galaxy cross-correlation amplitude for the multipole $l$  
can be estimated as \cite{Nakamura:2018oyy,Kable:2021yws}
\be
C_{l}^{\rm IG} \simeq \frac{3H_0^2 \Omega_{m}^{(0)}}
{D_0^2 l_{12}^2} \int_{0}^{z_*} {\rm d}z\,e^{-\tau} 
b_s H D^2 \Sigma\,{\cal F}\,P_m\left( \frac{l_{12}}{d_c} \right)\,,
\ee
where $l_{12}=l+1/2$, $\Omega_m^{(0)}$ is the present-day 
value of $\Omega_m$, $z_*$ denotes the redshift at recombination, 
$\tau$ is the visibility function, and $P_m(k)$ is today's matter 
power spectrum.
The function ${\cal F}$ is defined as
\be
{\cal F}=1-\frac{D'(N)}{D(N)}-\frac{\Sigma'(N)}{\Sigma(N)}\,,
\label{calF}
\ee
where $N=\ln a$.
If ${\cal F} > 0$ for redshifts $0 < z < z_*$, the cross-correlation 
$C_l^{\rm IG}$ is positive. 
Conversely, if ${\cal F} < 0$ over some redshift interval, 
$C_l^{\rm IG}$ can be negative. 
Observational results from the 2MASS 
and SDSS catalogs generally indicate positive 
ISW-galaxy cross-correlations \cite{Giannantonio:2008zi,Giannantonio:2012aa}.

In the $\Lambda$CDM model, we have $\Sigma = 1$, 
so Eq.~(\ref{calF}) reduces to ${\cal F} = 1 - D'(N)/D(N)$. 
In this case, the growth function $D$ at low redshifts 
is related to $\Omega_m$ through
$D'(N)/D(N) = (\Omega_m)^{\gamma}$, with 
$\gamma \simeq 0.55$ \cite{Wang:1998gt}. 
Since $0 < \Omega_m < 1$, the function $\mathcal{F}$ 
is always positive, implying $C_l^{\rm IG} > 0$.
In contrast, in our model, $\Sigma$ increases at low redshifts. 
Therefore, $\mathcal{F}$ may become negative, particularly 
when $\Sigma$ varies rapidly. 
This occurs in scalar Galileon DE models, where 
the negativity of $C_l^{\rm IG}$ is incompatible with observational data 
\cite{Kimura:2011td,Barreira:2012kk,Renk:2017rzu}. 
Such a problem can be alleviated in the modified scalar-tensor model
with a scalar potential proposed in Ref.~\cite{Tsujikawa:2025wca}.
In our DE model based on scalar-vector-tensor theories,
it is also possible to avoid negative ISW-galaxy cross-correlations
owing to the presence of the vector degree of freedom,
in addition to the scalar potential. 

In the right panel of Fig.~\ref{fig3}, we show the evolution of $\mathcal{F}$ 
corresponding to case (c) of Fig.~\ref{fig1} for four different values of $\nu_v$. 
For $\nu_v \lesssim 1$, we find that $\mathcal{F}$ remains positive at all redshifts. 
When $\nu_v = 10$, $\mathcal{F}$ becomes temporarily negative in the range 
$-0.05 < z < 0.52$, while remaining positive at other redshifts.  
By computing ${\cal F}$ in cases (a) and (b), we find that ${\cal F} > 0$ at all redshifts even for $\nu_v \lesssim {\cal O}(10)$. While $\mathcal{F}$ can be negative over certain redshift intervals for sufficiently large $s$, such as $s \gtrsim 0.3$, positive values of $\mathcal{F}$ can be realized at all $z$ for $s$
closer to 0. 
Compared to GP theories without a scalar potential studied in Ref.~\cite{Nakamura:2018oyy}, a wider range of $\nu_v$ 
gives rise to positive ISW-galaxy cross-correlations.
This behavior is attributed to the fact that, for fixed model 
parameters, $\Omega_\chi$ appearing in Eq.~(\ref{muana}) 
can be smaller than the total DE density due to the contribution 
of the scalar-field energy density at low redshifts. 
The evolution of $\mathcal{F}$ depends not only on $\nu_v$ 
but also on $s$ and $p$. 
Numerically, we find that when $\nu_v \lesssim 1$ and the phantom-divide 
crossing occurs, $\mathcal{F}$ remains positive in most cases, indicating that 
the model can be compatible with ISW-galaxy 
cross-correlation data \cite{Giannantonio:2008zi,Giannantonio:2012aa}. 

%%%%%%%%%%%%%%%%%%%%%%%%%%%%
\section{Conclusions}
\label{consec}
%%%%%%%%%%%%%%%%%%%%%%%%%%%%

Within the framework of scalar-vector-tensor theories, 
we have constructed a DE model with an explicit Lagrangian 
that allows for the phantom-divide crossing without introducing 
theoretical pathologies.
The model is characterized by a subclass of GP theories 
in the presence of a canonical scalar field with a potential.
In DE models based on GP theories with a luminal speed
of gravitational waves, the DE equation of state must lie
in the range $w_{\rm DE} < -1$ to avoid ghost instabilities. 
This no-go theorem for realizing the phantom-divide crossing
can be circumvented by the presence of a scalar potential
$V(\phi)$ that breaks shift symmetry.  
Our model is described by the action~(\ref{action})
with the coupling functions given in~(\ref{G23}).
For simplicity, we have chosen the exponential potential  
$V(\phi) = V_0 e^{-\lambda \phi / M_{\rm pl}}$, 
but it is also possible to consider other potentials that break
shift symmetry to realize the phantom-divide crossing.

Taking into account nonrelativistic matter and radiation,
we showed in Sec.~\ref{backsec} that the background equations
can be cast into the form of a dynamical system given by
Eqs.~(\ref{auto1})-(\ref{auto4}).
The DE equation of state can be expressed as in Eq.~(\ref{wde}),
showing that the deviation of $w_{\rm DE}$ from $-1$ arises
from the vector-field density parameter $\Omega_\chi$ and
the scalar-field dimensionless kinetic term $x^2$. 
While the vector field can initially drive the DE equation of state to
$w_{\rm DE} < -1$, the time variation of $\phi$ at late times
allows for the possibility of crossing $w_{\rm DE} = -1$ toward
the region $w_{\rm DE} > -1$. 

In Sec.~\ref{linearsec}, we derived the linear stability conditions
by computing the second-order actions for tensor, vector, and scalar
perturbations. The propagation speeds of the two transverse tensor modes
are equal to the speed of light, and no ghost instabilities arise.
The model is therefore compatible with the observational bounds
on the speed of gravity inferred from the GW170817 event and its
electromagnetic counterpart. 
The transverse vector modes are also free from
ghost and Laplacian instabilities. 
To avoid ghost and strong-coupling problems in the scalar sector, 
the parameter $s = p_2/p$ must be in the range $0 < s \leq 1/p$. 
If we further require the absence of a divergence in the squared
sound speed $c_\psi^2$ of the longitudinal scalar mode in the
asymptotic past, the additional bound $p(s+1) \geq 1$ 
needs to be satisfied.
In summary, the theoretically consistent parameter space is 
characterized by Eq.~(\ref{sp}).
 
In Sec.~\ref{dividesec}, we studied in detail how the phantom-divide crossing 
can occur in our model. We first derived the fixed points of the background 
solutions and showed that the de Sitter point (C), realized by the dominance 
of the vector-field energy, corresponds to a stable attractor.
In the early matter era, where the scalar kinetic energy is suppressed 
relative to the vector-field energy and the scalar potential energy 
($x^2 \ll \Omega_{\chi}$ and $x^2 \ll y^2$), the DE equation of 
state is approximately given by 
$w_{\rm DE} \simeq -1 - s\,\Omega_{\chi}/(\Omega_{\chi} + y^2)$.
As seen in the left panel of Fig.~\ref{fig2}, we have $\Omega_{\chi} \ll y^2$ 
in the asymptotic past, so that $w_{\rm DE}$ is close to $-1$. 
As $\Omega_{\chi}$ increases relative to $y^2$, $w_{\rm DE}$ decreases 
in the region $w_{\rm DE} < -1$. 
To realize the phantom-divide crossing at low redshifts, we require that the time variation of the scalar field contributes to $w_{\rm DE}$ in such a way that the condition~(\ref{xcon}) is satisfied. 
In Fig.~\ref{fig1}, we see that $w_{\rm DE}$ reaches its minimum 
and then begins to increase with time.
After crossing $w_{\rm DE}=-1$, it reaches a maximum and 
subsequently approaches $w_{\rm DE} = -1$ in the future. 
The deviation of $w_{\rm DE}$ before and after crossing $w_{\rm DE}=-1$ 
depends on the two model parameters $s$ and $\lambda$.
Both $s$ and $\lambda$ are required to be nonvanishing to 
achieve the evolution from $w_{\rm DE} < -1$ to $w_{\rm DE} > -1$.
As seen in the right panel of Fig.~\ref{fig2}, there are neither 
ghost nor Laplacian instabilities throughout the cosmological 
evolution from the radiation era.

In Sec.~\ref{growthsec}, we studied the evolution of linear perturbations by considering the density contrast $\delta_m$ of nonrelativistic matter, as well as the gauge-invariant gravitational potentials $\Psi$ and $\Phi$. Under the quasi-static approximation for modes deep inside the Hubble radius, we showed that the two dimensionless quantities $\mu$ and $\Sigma$, defined through Eqs.~(\ref{Psia}) and (\ref{Psia2}), are equivalent to each other and take the form given in Eq.~(\ref{muana}).
The conditions $s>0$ and $c_{\psi}^2>0$ are required to avoid ghost 
and Laplacian instabilities. As a result, the gravitational interaction for 
perturbations relevant to large-scale structure is stronger than in the $\Lambda$CDM model, i.e., $\mu = \Sigma > 1$.
Since $c_{\psi}^2$ is influenced by the presence of intrinsic vector 
modes through the quantity $\nu_v$, there is some freedom in 
realizing values of $\mu$ and $\Sigma$ close to 1.
Moreover, the presence of the scalar potential $V(\phi)$ can 
reduce the value of $\Omega_{\chi}$ at low redshifts 
compared to GP theories with $V(\phi)=0$, 
resulting in reduced deviations of $\mu$ and $\Sigma$ from 1. 
As seen in the left panel of Fig.~\ref{fig3}, the deviation of 
$\mu$ from 1 is suppressed for $\nu_v \ll 1$. 
Furthermore, the quantity ${\cal F}(z)$ associated with the 
ISW-galaxy cross-correlation can be positive at all redshifts 
over a broad region of the model parameter space.

We have thus shown that our model, constructed within the framework
of scalar-vector-tensor theories, can realize the phantom-divide crossing
without encountering theoretical pathologies such as ghosts, Laplacian
instabilities, or strong coupling.
At the background level, the model introduces two additional parameters,
$s$ and $\lambda$, beyond those of the $\Lambda$CDM model.
When the evolution of linear perturbations is included in the analysis,
two further parameters, $p$ and $\nu_v$, appear, affecting the quantities
$\mu$ and $\Sigma$.
A detailed investigation of the observational constraints on the four 
parameters $s$, $\lambda$, $p$, and $\nu_v$ is left for future work.

%%%%%%%%%%%%%%%%%%%%%%%%%%%
\section*{Acknowledgements}
%%%%%%%%%%%%%%%%%%%%%%%%%%%%

The author thanks JSPS KAKENHI Grant No.~22K03642 and 
Waseda University Special Research Projects 
(Nos.~2025C-488 and 2025R-028) for their support.

\bibliographystyle{mybibstyle}
\bibliography{bib}

\end{document}